\documentclass{article}

\usepackage{PRIMEarxiv}

\usepackage[utf8]{inputenc} %
\usepackage[T1]{fontenc}    %
\usepackage{hyperref}       %
\usepackage{url}            %
\usepackage{booktabs}       %
\usepackage{amsfonts}       %
\usepackage{nicefrac}       %
\usepackage{microtype}      %
\usepackage{lipsum}
\usepackage{fancyhdr}       %
\usepackage{graphicx}       %
\usepackage{subcaption}
\usepackage[font=small, labelfont=bf]{caption}       %
\graphicspath{{figures/}}     %
\usepackage{epstopdf}
\usepackage[table,xcdraw]{xcolor}
\usepackage[table]{xcolor}  %
\usepackage{amsmath}  %
\usepackage[numbers,sort&compress]{natbib}

\usepackage{multirow}
\usepackage{placeins}
\usepackage{soul,color}

\title{Structure-based out-of-distribution (OOD) materials property prediction: a benchmark study
\thanks{\textit{\underline{Citation}}: 
\textbf{Omee et al.. Structure-based OOD materials property prediction benchmark study. 15 Pages......DOI:000000/11111.}} 
}

\author{
Sadman Sadeed Omee\\
 Department of Computer Science and Engineering\\
  University of South Carolina\\
  Columbia, SC 29201   
   \And
 Nihang Fu\\
 Department of Computer Science and Engineering\\
  University of South Carolina\\
  Columbia, SC 29201 \\  
  \And
Rongzhi Dong\\
 Department of Computer Science and Engineering\\
  University of South Carolina\\
  Columbia, SC 29201 \\  
  \And 
 Ming Hu *\\
 Department of Mechanical Engineering\\
  University of South Carolina\\
  Columbia, SC 29201 \\
  \texttt{hu@sc.edu} \\
 \And
 Jianjun Hu *\\
 Department of Computer Science and Engineering\\
  University of South Carolina\\
  Columbia, SC 29201 \\
  \texttt{jianjunh@cse.sc.edu} \\
}

\begin{document}
\maketitle

\begin{abstract}
In real-world material research, machine learning (ML) models are usually expected to predict and discover novel exceptional materials that deviate from the known materials. It is thus a pressing question to provide an objective evaluation of ML model performances in property prediction of out-of-distribution (OOD) materials that are different from the training set distribution. 
Traditional performance evaluation of materials property prediction models through random splitting of the dataset frequently results in artificially high performance assessments due to the inherent redundancy of typical material datasets. Here we present a comprehensive benchmark study of structure-based graph neural networks (GNNs) for extrapolative OOD materials property prediction. We formulate five different categories of OOD ML problems for three benchmark datasets from the MatBench study. Our extensive experiments show that current state-of-the-art GNN algorithms significantly underperform for the OOD property prediction tasks on average compared to their baselines in the MatBench study, demonstrating a crucial generalization gap in realistic material prediction tasks. We further examine the latent physical spaces of these GNN models and identify the sources of CGCNN, ALIGNN, and DeeperGATGNN's significantly more robust OOD performance than those of the current best models in the MatBench study (coGN and coNGN), and provide insights to improve their performance.

\end{abstract}


\section{Introduction}

Machine learning (ML)-based models have swiftly emerged as the state-of-the-art (SOTA) performers in a wide range of materials informatics problems such as materials property prediction~\cite{cgcnn,megnet,alignn,deepergatgnn,matformer,crabnet,roost}, crystal structure prediction~\cite{gnoa,paretocsp,hu2023deep,qi2023latent,wang2023magus}, material generation \cite{gnome,unimat,cubicgan,zhao2023physics}, high-throughput screening\cite{fanourgakis2020universal,ojih2023screening}, and inverse design of materials~\cite{,han2023design}. Among these, one of the most exciting applications of ML-based models is to predict various properties of materials given their compositions, or structures. Composition-based ML models have shown limited prediction performance \cite{seko2017representation,ward2016general,ward2018matminer}, as most material properties are highly dependent on their crystal structures. Recent research has demonstrated that structure-based deep learning (DL) models can achieve significantly better accuracy in predicting materials properties compared to methods that exclusively rely on composition descriptors~\cite{fung2021benchmarking,materialsatlas,dunn2020benchmarking}. Especially, graph neural network (GNN) models have been widely utilized for this purpose due to their demonstrated superior effectiveness in this task~\cite{alignn,deepergatgnn,dimenet,gasteiger2021gemnet,fung2021benchmarking}. This is because GNNs excel at capturing the local environment of each atom by considering its neighboring atoms and their interactions, which is crucial in determining the macro-properties of a material~\cite{reiser2022graph,louis2020graph, kong2022density,cong2023improving,xiao2024graph}. 

There are several benchmark studies conducted for evaluating the performances of existing ML methods. Dunn et al.~\cite{dunn2020benchmarking} presented the Matbench benchmark test suite and an automated procedure for evaluating ML models for predicting material properties. This benchmark contains nine distinct structure based property prediction tasks.
Remarkably, the SOTA GNN model coGN \cite{ruff2023connectivity} has consistently demonstrated superior performance with an MAE of 0.017 eV for formation energy prediction and 0.156 eV for bandgap prediction, while the top positions on the leaderboard~\cite{matbench_leaderboard} for all nine tasks are all secured by structure-based GNN models. However, the excellent performances of these GNN models are overestimated as verified by our work. 
We find that the reported superior performances of SOTA models in the MatBench study are originated from their performance evaluation method, in which an entire dataset is randomly split into the training and test sets leading to high similarity between both sets due to the high sample redundancy of materials databases \cite{matda}. This dataset redundancy in material databases such as ICSD~\cite{icsd}, Materials Project~\cite{materialsproject}, OQMD~\cite{kirklin2015open}, and AFLOW~\cite{curtarolo2012aflow}, is caused by the historical iterative tinkering process of experimental material discovery and accumulation, which tends to generate many materials with high similarity. 
Moreover, studies have revealed that current ML models have low generalization performance on material datasets for test samples with different data distributions, and their performances are frequently overestimated because of high dataset redundancy~\cite{li2023exploiting,li2023critical,xiong2020evaluating,meredig2018can,matda}. Li et al.~\cite{li2023critical} discovered that ML models trained on Materials Project 2018 data may experience a significant decline in performance when applied to new materials introduced in Materials Project 2021 data, primarily due to a shift in data distribution. Consequently, the resulting high prediction performance over these test sets in Matbench assumed homogeneity in terms of composition, structure, or properties and was randomly distributed within the entire dataset space. This material property performance evaluation approach, guided by the assumption of independent and identically distributed (i.i.d.) data proved inadequate in replicating their performance in real-world material discovery applications. In practical scenarios, ML models are often utilized to discover or screen outlier materials that deviate from the distribution of the training set and need their property to be predicted. Moreover, in real-world situations, researchers commonly focus on a limited number of outlier materials that fall outside the typical distribution, referred to as \textit{OOD materials}. These materials could be located in a chemical space with limited known counterparts, or they might display exceptionally high or low property values~\cite{matda}. An evaluation of ML-based material property prediction performance in these particular situations was not provided in the MatBench study and in the literature to our knowledge. 

Recently, within the domain of ML, researchers started to extensively investigate OOD generalization resulting from changes in data distribution between the original and target domains, primarily in the context of transfer learning ~\cite{wenzel2022assaying}, domain generalization~\cite{wang2022generalizing,shen2021towards}, causal learning~\cite{scholkopf2021toward}, and domain adaptation ~\cite{wilson2020survey}. This shift in distribution is a key concern in these areas. Most of these methods are still unexplored for improving the material property prediction performance in OOD materials. To our knowledge, there has not been a comprehensive benchmark study of ML models for OOD property prediction for inorganic materials.
Another shortcoming of current ML approaches for material property prediction is that the ML models are usually trained without considering the distribution of the test set. In practical materials property prediction tasks, the compositions or structures of the target materials are already available, which can be and should be used as guidance information for training better ML models for property prediction of these target materials \cite{matda}. Furthermore, Schrier et al.~\cite{schrier2023pursuit} highlighted that material scientists usually prioritize investigating the properties of new materials with unique compositions or characteristics, which frequently gives rise to challenging OOD ML problems. This underscores the critical need for a systematic investigation into the prediction problem of the properties of OOD materials. 

\begin{figure}[!htb]
    \centering
    \includegraphics[width=\textwidth]{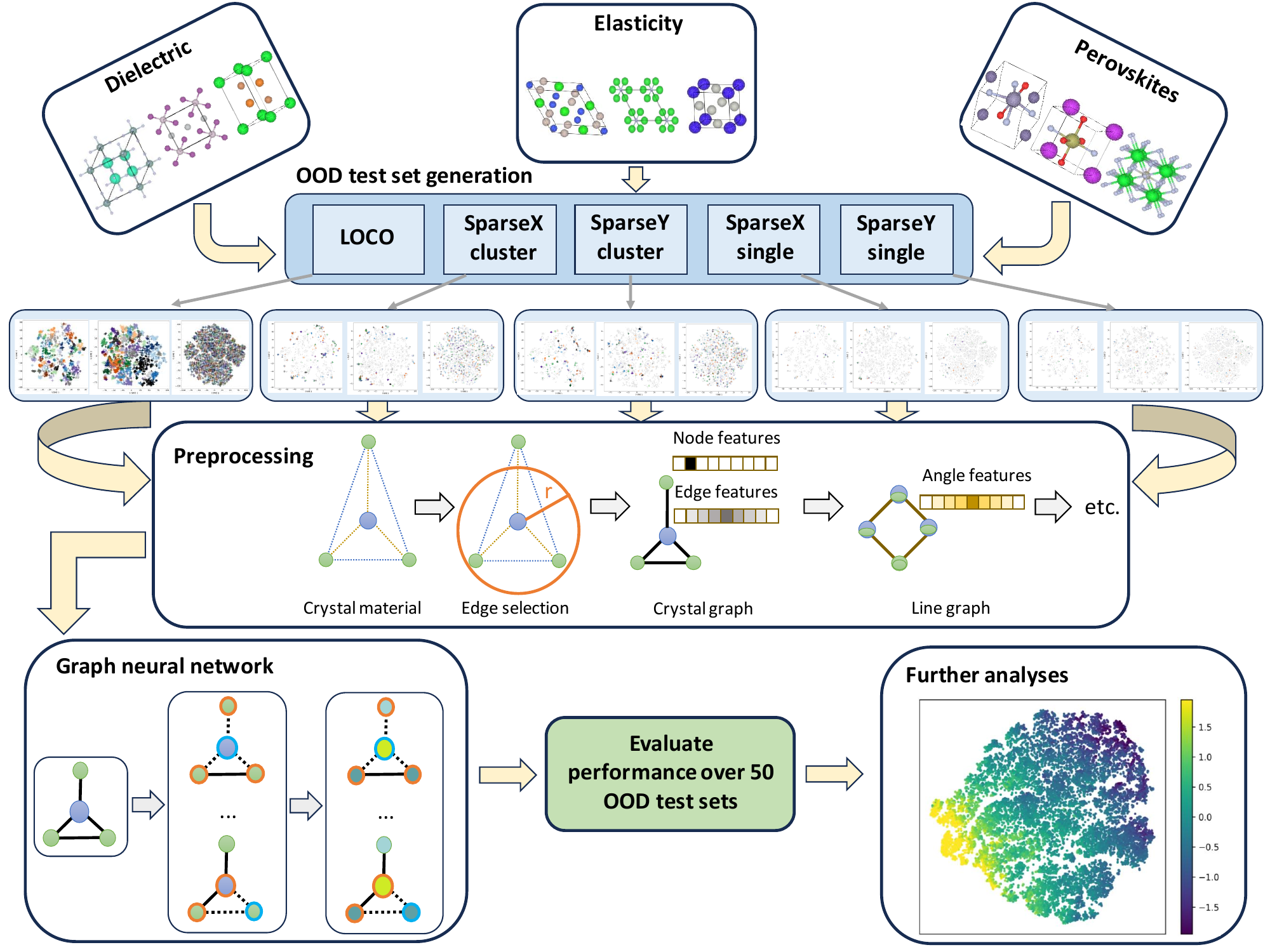}
    \caption{\textbf{The overall framework and workflow of our OOD materials benchmark.}\\ First, we generate OOD test sets for the three datasets chosen, where we propose five different methods to split each dataset into 50 folds, ensuring the test set varies in distribution from the training set in each fold. Next, we perform preprocessing steps such as input representation, data scaling, etc. for the GNNs. Subsequently, we train the GNN models and compile the test set results. After that, we evaluate the performance over the 50 folds for each OOD target generation method. We conduct additional analyses on the obtained results, including investigating the physical latent spaces of the GNN models to understand their characteristics in predicting properties of OOD materials.}
    \label{fig:framework}
\end{figure}

Due to the sheer dominance of GNNs in the MatBench study, this work aims for a benchmark study for GNN-based OOD materials property prediction. Our work is complementary to the benchmark study by Fung et al.~\cite{fung2021benchmarking} on GNNs performance on materials property prediction. However, their analysis did not consider OOD materials as test sets. Although several works have been done on the OOD topic in general~\cite{yang2022openood,gui2022good,shen2021towards,koh2021wilds} including a similar benchmark work~\cite{shimakawa2024extrapolative} for organic materials, to our knowledge, our work is the first OOD benchmark work of structure-based GNNs for inorganic materials. In particular, this work focuses on effectively predicting properties of minority or outlier material clusters that exhibit different distributions compared to the training set. These scenarios are characterized by the core issue of OOD ML. A general framework of our benchmark study is presented in Fig.~\ref{fig:framework}. Details about the datasets and the OOD target generation methods can be found in Section~\ref{subsec:models_data}%
Our contributions are summarized as follows:

\begin{itemize}
    \item We proposed a set of OOD material property prediction benchmark problems for three datasets from the MatBench study, where each category of OOD targets possesses unique characteristics, creating a more realistic and complex challenge for current SOTA GNN algorithms.
    \item Through comprehensive experiments on these OOD problems, we benchmarked GNN algorithms for property prediction for these OOD datasets. We showed that current GNN algorithms have limited generalization capabilities and are not well-suited for real-world OOD material property prediction tasks, with the exception of a few cases for CGCNN, ALIGNN, and DeeperGATGNN. In general, all the algorithms perform worse on average on the OOD test problems than their baseline performance in the original MatBench study, suggesting methods like domain adaptation are needed to improve their OOD prediction performance.
    \item By delving into the physical latent spaces of the GNN models, we identified possible reasons for the comparatively better performance of CGCNN, ALIGNN, and DeeperGATGNN and the subpar performance of current top models in the MatBench leaderboard - coGN and coNGN.
\end{itemize}

\section{Results}
\label{sec:headings}

\subsection{OOD benchmark problems, models and datasets} \label{subsec:models_data}

We analyzed eight GNN models for material property regression tasks using three datasets sourced from MatBench~\cite{dunn2020benchmarking} and mentioned in Table~\ref{tab:models}. Details about the GNN models can be found in Section~\ref{subsec:sota_gnns}.
The raw dataset details are shown in Table~\ref{tab:datasets}. For simplicity, we refer to the matbench\_dielectric dataset as the `dielectric dataset', the matbench\_log\_gvrh dataset as the `elasticity dataset', and the matbench\_perovskites dataset as the `perovskites dataset'.

\begin{table}[!htb]
\caption{List of the GNN models used in this work.}
\centering
\label{tab:models}
\begin{tabular}{l c c}
\hline
\rowcolor[HTML]{C0C0C0}
\textbf{GNN model} & \textbf{Source} & Publishing year\\ \hline
CGCNN & Xie and Grossman~\cite{cgcnn} & 2018 \\ \hline
SchNet & Schütt et al.~\cite{schnet} & 2018 \\ \hline
MEGNet & Chen et al.~\cite{megnet} & 2019 \\ \hline
DimeNet++ & Gasteiger et al.~\cite{dimenet++} & 2020 \\ \hline
ALIGNN & Choudhary and DeCost~\cite{alignn} & 2021 \\ \hline
DeeperGATGNN & Omee et al.~\cite{deepergatgnn} & 2022 \\ \hline
coGN & Ruff et al.~\cite{cogn} & 2023 \\ \hline
coNGN & Ruff et al.~\cite{cogn} & 2023 \\ \hline
\end{tabular}
\end{table}

\begin{table}[!htb]
\caption{Details of the three benchmark datasets used in this work.}
\centering
\label{tab:datasets}
\begin{tabular}{l c c c c c}
\hline
\rowcolor[HTML]{C0C0C0}
\textbf{Dataset} & \begin{tabular}[c]{@{}c@{}}\textbf{Target}\\ \textbf{property}\end{tabular} & \begin{tabular}[c]{@{}c@{}}\textbf{Total}\\ \textbf{samples}\end{tabular} & \begin{tabular}[c]{@{}c@{}}\textbf{Original}\\ \textbf{source}\end{tabular} & \begin{tabular}[c]{@{}c@{}}\textbf{MatBench best}\\\textbf{algorithm (MAE)}\end{tabular} & \textbf{Unit}\\ \hline
matbench\_dielectric             & \begin{tabular}[c]{@{}c@{}}Refractive\\index\end{tabular} & 4764  & \begin{tabular}[c]{@{}c@{}}Materials\\Project~\cite{materialsproject,petousis2017high}\end{tabular}                                                 & MODNet~\cite{modnet} (0.2711)  & Unitless \\ \hline
matbench\_log\_gvrh           & \begin{tabular}[c]{@{}c@{}}Shear\\modulus\end{tabular}                    & 10987    & \begin{tabular}[c]{@{}c@{}}Materials\\Project~\cite{materialsproject,petousis2017high}\end{tabular}                                              & coNGN~\cite{cogn} (0.0670) & $\log$10(GPa)               \\ \hline
matbench\_perovskites            & \begin{tabular}[c]{@{}c@{}}Formation\\energy\end{tabular} &  18928 & Castelli et al.~\cite{castelli2012new} & coGN~\cite{cogn} (0.0269)       & eV/unit cell         \\ \hline
\end{tabular}
\end{table}

In realistic scenarios, the test set often deviates significantly from the training and validation set in terms of distribution. Rather than applying conventional methods like random train-test splitting or $k$-fold cross-validation (which also relies on random splitting), we proposed five practical scenarios for predicting material properties.
These scenarios are designed to focus on properties of less common or underrepresented materials in the dataset, which are often of particular interest to material researchers who are more interested in discovering novel exceptional outlier materials. For each raw dataset, we outlined five methods %
(See Section \ref{subsec:test_generation} for details) for selecting which samples from the sparse property or structure space will be designated as the target test samples. Overall, each target generation method generates 50 clusters, where each cluster has a different distribution from the others. For each fold, we selected a cluster as the test set, and the rest as training and validation sets, and averaged the results over 50 folds to get the final result.

\subsubsection{OOD test set generation} \label{subsec:ood_generation}
\label{subsec:test_generation}

\begin{figure}[ht!] 
    \begin{subfigure}[t]{0.33\textwidth}
        \includegraphics[width=0.99\textwidth]{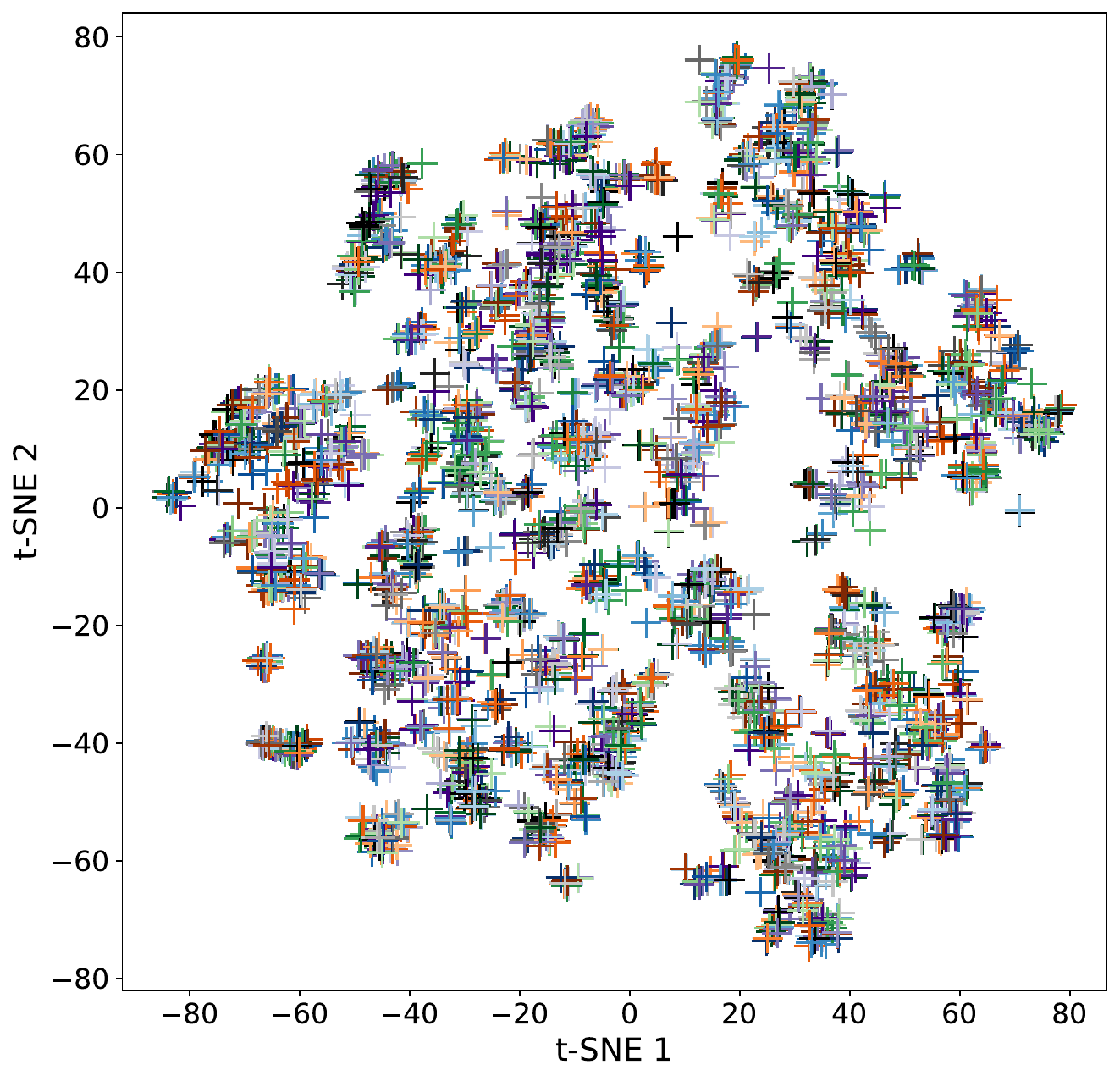}
        \caption{50-fold CV with Random splitting}
        \vspace{3pt}
        \label{fig:whole}
    \end{subfigure}
 \begin{subfigure}[t]{0.33\textwidth}
        \includegraphics[width=0.99\textwidth]{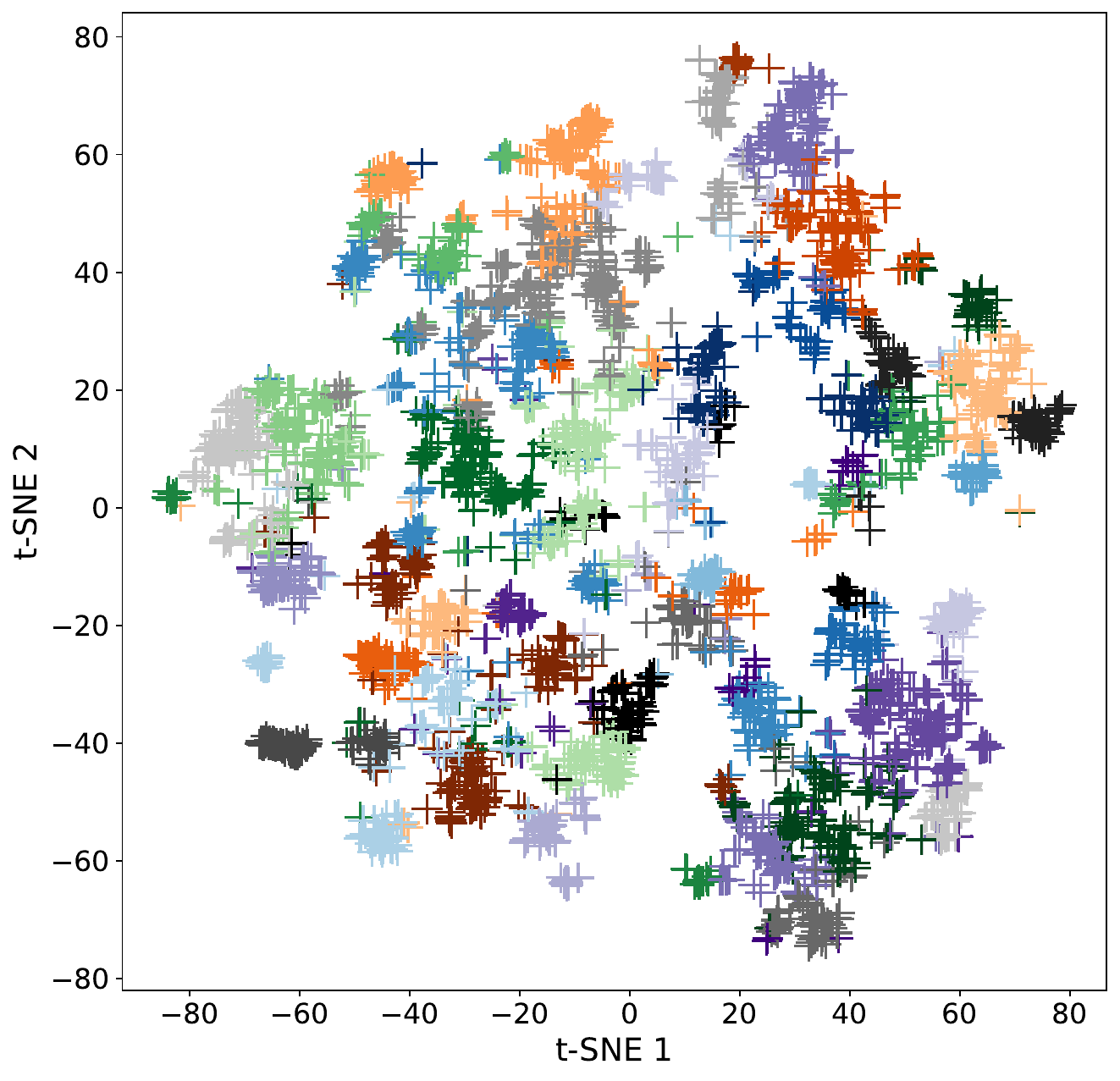}
        \caption{LOCO}
        \vspace{-3pt}
        \label{fig:loco}
    \end{subfigure}    
    \begin{subfigure}[t]{0.33\textwidth}
        \includegraphics[width=0.99\textwidth]{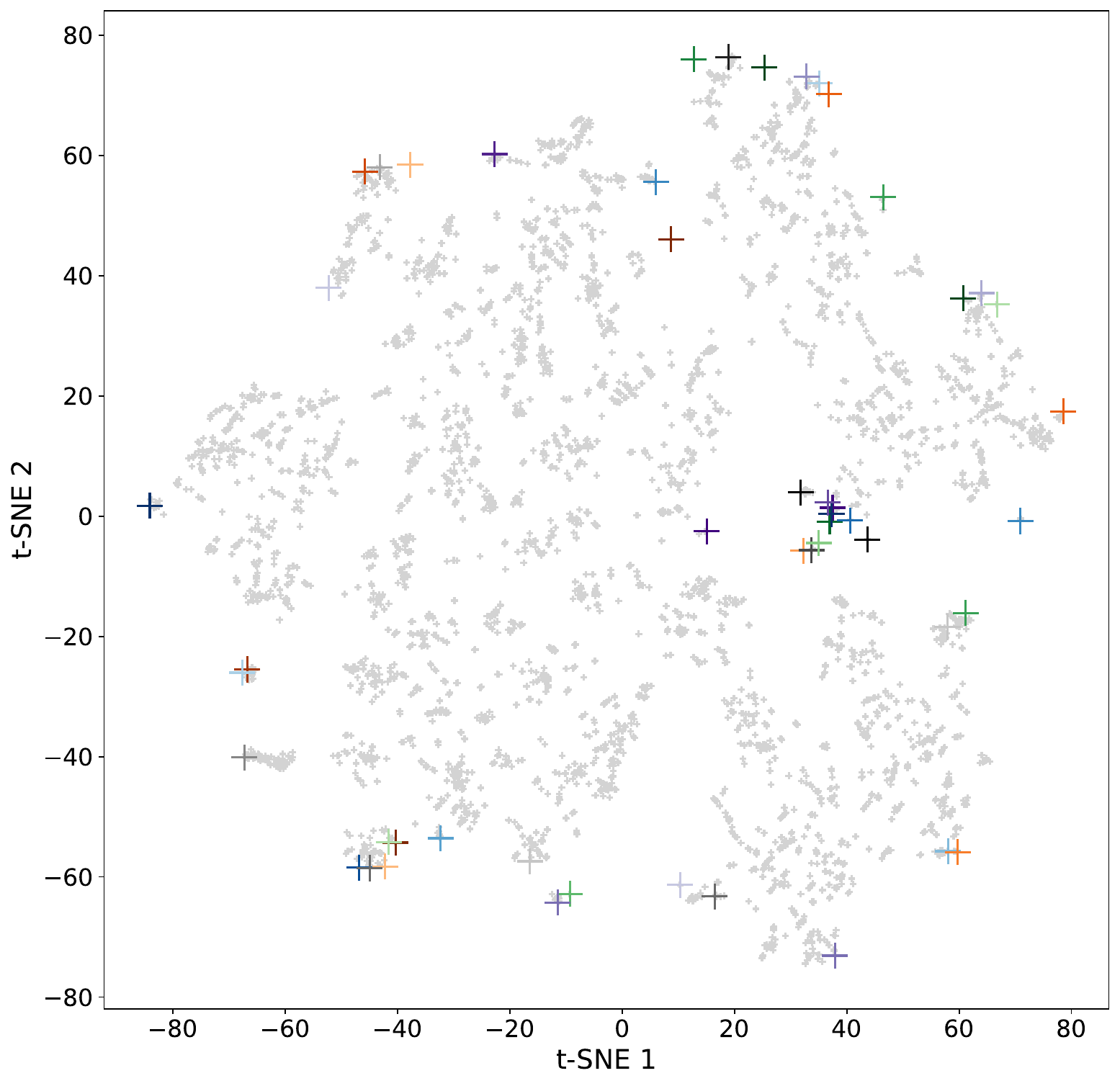}
        \caption{SparseXsingle samples}
        \vspace{3pt}
        \label{fig:sparse_x_single}
    \end{subfigure} 
 \begin{subfigure}[t]{0.33\textwidth}
        \includegraphics[width=0.99\textwidth]{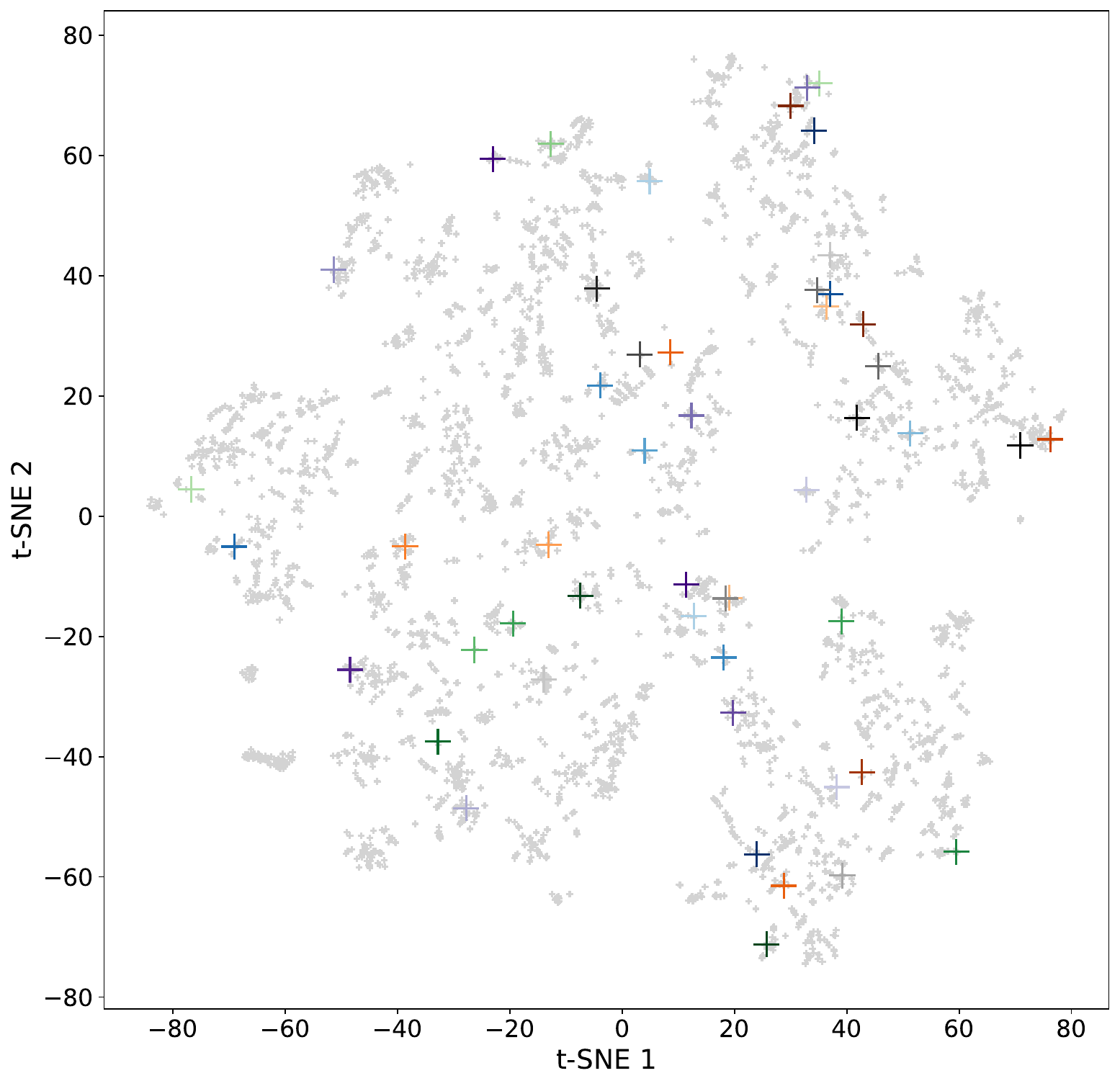}
        \caption{SparseYsingle samples}
        \vspace{3pt}
        \label{fig:sparse_y_single}
    \end{subfigure}              
    \begin{subfigure}[t]{0.33\textwidth}
        \includegraphics[width=0.99\textwidth]{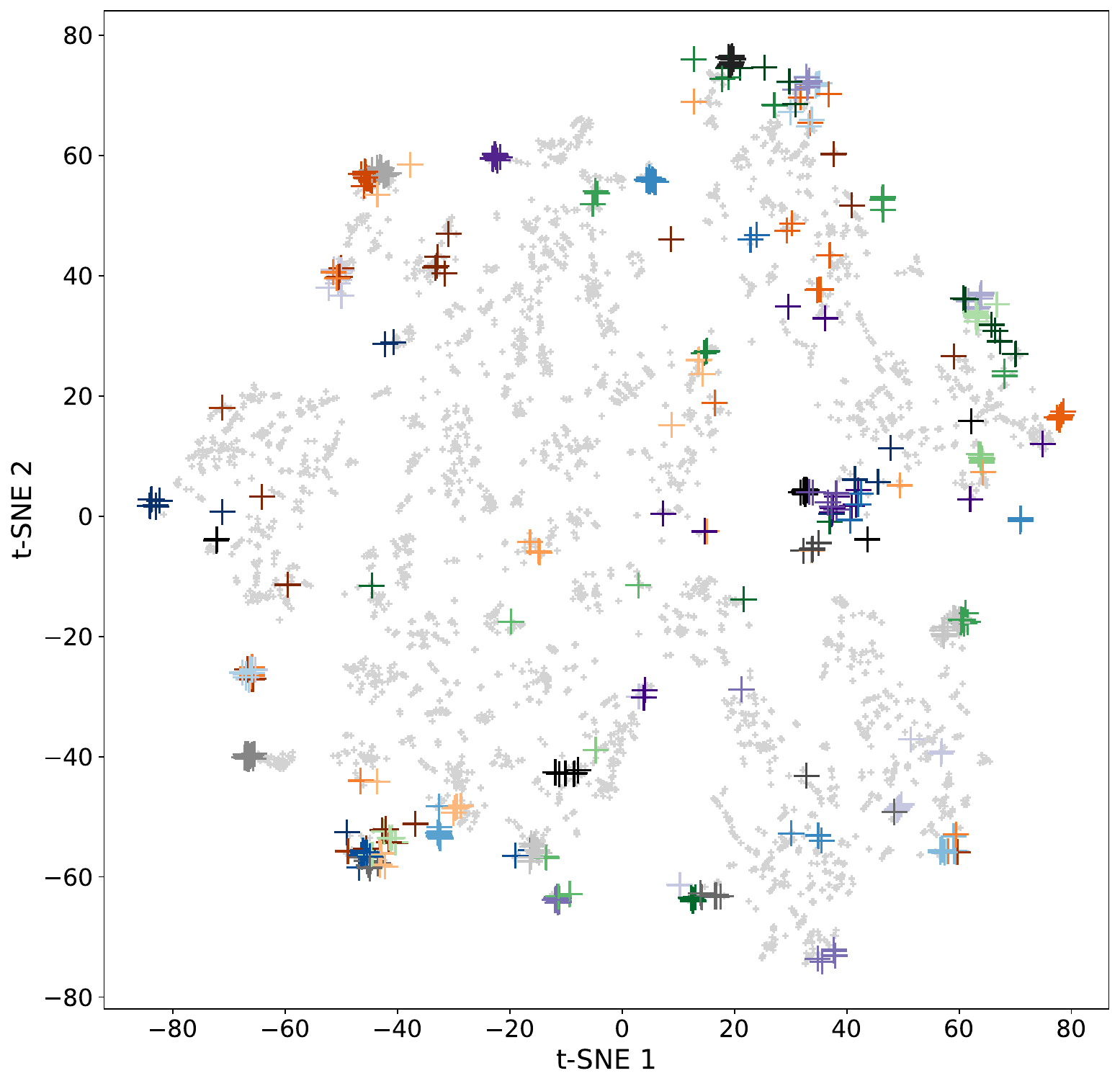}
        \caption{SparseXcluster samples}
        \vspace{-3pt}
        \label{fig:sparse_x_cluster}
    \end{subfigure}
    \begin{subfigure}[t]{0.33\textwidth}
        \includegraphics[width=0.99\textwidth]{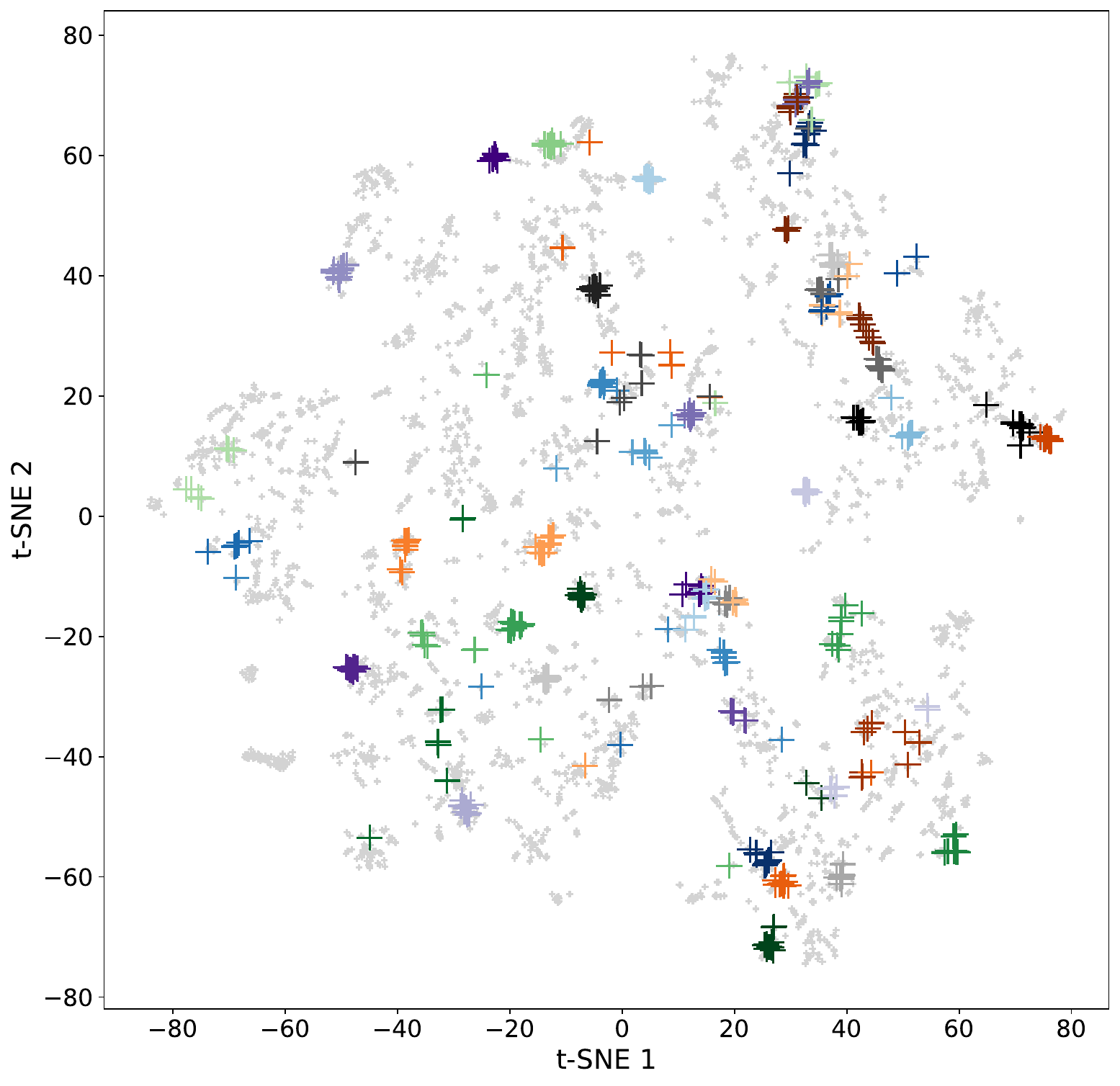}
        \caption{SparseYcluster samples}
        \vspace{-3pt}
        \label{fig:sparse_y_cluster}
    \end{subfigure} 

   \caption{\textbf{Distribution of standard cross-validation (CV) test set and five OOD test sets using various target generation methods for the dielectric dataset.}\\ (a) 50-fold CV (with random splitting) of the whole dielectric dataset with 4,764 samples represented by cross symbols with 50 different colors. (b) Leave-one-cluster-out target (LOCO) clusters. (c) In SparseXsingle, 50 test samples are represented by cross symbols with 50 different colors, and grey points represent the remaining samples. (d) In SparseYsingle, 50 test samples are represented by cross symbols with 50 different colors, and grey points represent the remaining samples. (e) SparseXcluster displays 50 test clusters represented by cross symbols with 50 different colors, and grey points represent the remaining samples. (f) SparseYcluster displays 50 test clusters represented by cross symbols with 50 different colors, and grey points represent the remaining samples.}
  \label{fig:distribution}
\end{figure}

In typical real-world scenarios, researchers are acquainted with their target materials of interest, often lacking labeled samples. In this work, we specifically concentrate on instances where the target set comprises no labeled samples. Accordingly, we propose the following target set generation methods to simulate real-world conditions for materials property prediction.

\paragraph{Leave-one-cluster-out (LOCO)}
Meredig et al. \cite{meredig2018can} proposed this approach in their assessment of the generalization performance of machine learning models for predicting material properties. 
Initially, we apply the $k$-means algorithm~\cite{lloyd1982least} based on the orbital-field matrix (OFM) features~\cite{pham2017machine} to cluster the whole dataset into 50 clusters. Subsequently, we evaluate the models' performance by iteratively using each of the clusters as test sets. 
While enhancing the widely employed random splitting method to mitigate performance overestimation, it still incorporates all samples, especially those located in densely redundant areas. This implies that it retains a susceptibility to some degree of overestimation.

\paragraph{Single-point targets with the lowest structure density (SparseXsingle)}

In this method, we begin by converting material structures into the 1024-dimension OFM feature space. Subsequently, we apply the t-distributed stochastic neighbor embedding (t-SNE) \cite{van2008visualizing} for dimension reduction converting the OFM feature space to a 2D space ($x$-value). Following this, we calculate the density for each data point in the 2D space and select 500 samples with the lowest density. We apply $k$-means clustering on these chosen samples to convert them into 50 clusters. Finally, we extract one sample from each cluster, yielding a test set with 50 target samples. 

\paragraph{Single-point targets with the lowest property density (SparseYsingle)}

In this method, we follow the preprocessing method of SparseXsingle, where all structures are converted into 1024-dimension OFM features. Following this, we sort the samples based on their property values ($y$-value). This process estimates the density of $y$-values using kernel density estimation for each data point and picks the 500 samples with the lowest density. Then we apply the $k$-means clustering to convert these chosen samples to 50 clusters. From each cluster, we pick one sample, obtaining our test set with 50 target samples.

\paragraph{Cluster targets with the lowest structure density (SparseXcluster)}

This sparse cluster target set generation method is similar to the SparseXsingle method. However, after $k$-means clustering, rather than selecting just one sample, we extend the selection to include $N$ nearest neighbors for each chosen sample to form the target cluster. The process of picking neighbors ensures that no sample is selected into multiple target clusters with the neighbors determined by the Euclidean distance of OFM features.

\paragraph{Cluster targets with the lowest property density (SparseYcluster)}

This sparse cluster target set generation method closely resembles the SparseYsingle method but with a notable distinction. Following $k$-means clustering, instead of selecting a single sample, we expand the selection to include $N$ nearest neighbors for each chosen sample to create a target cluster. The neighbor-picking process is conducted to prevent any sample from being selected into multiple target clusters. The determination of neighbors is based on the Euclidean distance of OFM features.

The distribution of the whole dielectric datasets and their different target sets are shown in Fig.~\ref{fig:distribution}. Additionally, Supplementary Fig. S1 and S2 provide visualizations for the elasticity and perovskites datasets. It can be observed that realistic target sets predominantly reside in sparser regions, whereas commonly employed random splitting tends to align with dense areas exhibiting a distribution akin to that of the training set. In total, we prepared 3 datasets, the dielectric dataset, the elasticity dataset, and the perovskites dataset, for the benchmark evaluations, and each of them contains LOCO, SparseXsingle, SparseXcluster, SparseYsingle, and SparseYcluster test sets for regression. The number of samples for each cluster of these datasets is shown in Supplementary Tables S1, S2, and S3.

\subsection{Performance comparison on OOD test sets}
\label{subsec:comp_ood}

Here we report the OOD performance of selected GNN models for three datasets. The training hyperparameters used for this benchmark study are listed in the Supplementary file. The results on the dielectric dataset for five different OOD target generation methods are summarized in Table~\ref{table:dielectric}. For the LOCO generation method, we find that CGCNN achieved the SOTA OOD test results on the dielectric dataset (MAE: 0.5144), and DeeperGATGNN performed the second best, showing a 14.91\% increase in MAE (0.5911). The other GNN models performed significantly worse than these two for the LOCO targets, with DimeNet++ registering the highest MAE at 2.7720. This discrepancy in performance can be attributed to the fact that in contrast to random train-test splitting or cross-validation, the LOCO targets tend to have different distributions compared to the training sets (refer to Fig.~\ref{fig:distribution}f). This introduces increased complexity and challenges for conventional ML/DL models such as MEGNet that are well-trained to achieve good prediction performance on i.i.d. test sets. The OOD test sets for the SparseXcluster and SparseYcluster datasets are formed through a two-step process. Initially, 50 seed samples with the highest sparsity in the OFM space are chosen. From these, 10 samples that are most similar to the seed samples (depending on the $x$-axis or $y$-axis) are selected. The main goal for these target generation methods is to evaluate the effectiveness of an ML/DL algorithm to predict the properties without using closest neighbors. For the SparseXcluster targets, CGCNN achieved the best performance on both the dielectric dataset (MAE: 0.6006), followed by ALIGNN with an MAE increase of 15.77\% to 0.6953. However, for the SparseYcluster targets, DeeperGATGNN achieved the SOTA MAE of 0.3959, which is  10.10\% less than that of its closest model ALIGNN (MAE: 0.4359). The remaining models again displayed significantly subpar performance compared to these three models for SparseXcluster and SparseYcluster targets.
The single-point sparse X and sparse Y test sets are distinctive because they consist of only one sample each, with all other samples being utilized for training and validation. For these two OOD test sets, CGCNN outperformed all other models for the SparseXsingle targets (MAE: 0.9888), with a staggering 52.86\% decrease in MAE than the second best performing model, ALIGNN (MAE: 1.5115). For the SparseYsingle targets, ALIGNN achieved the lowest MAE (0.2513), which is slightly better (8.75\%) than its closest performer, DeeperGATGNN (MAE: 0.2733). In contrast, other models were consistently outperformed by these three models by a large margin for the single-point Sparse X and Y targets, with SchNet achieving the highest MAE (3.9767) for SparseXsingle targets, and DimeNet++ obtaining the highest MAE (2.5866) for SparseYsingle targets.

Table~\ref{table:elasticity} shows the summarized results for five types of OOD test sets on the elasticity dataset. CGCNN achieved the SOTA MAE for both LOCO (0.0585 $\log_{10}$(GPa)) and SparseXcluster (0.0499 $\log_{10}$(GPa)) targets, which is significantly better than the second best performing model ALIGNN (MAE: 0.0974 $\log_{10}$(GPa), and 0.0834 $\log_{10}$(GPa), respectively) for both these OOD targets (66.50\%, and 67.13\%, respectively). On the other hand, ALIGNN achieved the SOTA MAEs on the rest of the OOD targets (SparseYcluster: 0.0631 $\log_{10}$(GPa), SparseXsingle: 0.0853 $\log_{10}$(GPa), and SparseYsingle: 0.0450 $\log_{10}$(GPa)). CGCNN performed the second best for the SparseYcluster and SparseXsingle targets (MAE: 0.0752 $\log_{10}$(GPa), and 0.0895 $\log_{10}$(GPa), respectively), with an increase in MAE of 16.09\%, and 4.92\%, respectively, while DeeperGATGNN performed the second best for the SparseYsingle targets (MAE: 0.0807 $\log_{10}$(GPa)) with a remarkable 79.33\% increase in MAE. Other models performed consistently worse than CGCNN, ALIGNN, and DeeperGATGNN on the elasticity dataset, with MEGNet registering the worst MAE for the LOCO (1.4468 $\log_{10}$(GPa)), SparseXcluster (1.4113 $\log_{10}$(GPa)), and SparseYcluster (1.5659 $\log_{10}$(GPa)) targets, DimeNet++ achieving the worst MAE for the SparseXsingle targets (1.3214 $\log_{10}$(GPa)), and SchNet recording the poorest MAE for the SparseYsingle targets (1.4855 $\log_{10}$(GPa)).

Results on the perovskites dataset are summarized in Table~\ref{table:perovskites}. We can find that DeeperGATGNN outperformed all other algorithms for four out of five OOD targets (MAEs - LOCO: 0.036 eV/unit cell, SparseXcluster: 0.0464 eV/unit cell, SparseYcluster: 0.0333 eV/unit cell, SparseXsingle: 0.0373 eV/unit cell), demonstrating superior performance on the perovskites data. ALIGNN trailed its performance with a significant increase in MAE of 5.75\%, 0.86\%, 22.52\%, and 2.40\%, for the LOCO (MAE: 0.0386 eV/unit cell), SparseXcluster (MAE: 0.0468 eV/unit cell), SparseYcluster (MAE: 0.0341 eV/unit cell), and SparseXsingle (MAE: 0.0457 eV/unit cell) targets, respectively. ALIGNN achieved the SOTA performance for the single-point Sparse Y targets (MAE: 0.0243 eV/unit cell), outperforming DeeperGATGNN (MAE: 0.0259 eV/unit cell) by a slight margin (6.58\%). All other models demonstrated their low performance on the OOD perovskites data, with DimeNet++ achieving the highest MAEs for LOCO (MAE: 1.4666 eV/unit cell), SparseXcluster (MAE: 1.4567 eV/unit cell), and SparseXsingle targets (MAE: 1.5248 eV/unit cell), and SchNet registering poorest MAEs for SparseYcluster (MAE: 1.4736 eV/unit cell), and SparseYsingle (MAE: 1.5173 eV/unit cell) targets.

While the latest SOTA GNN models try to outperform each other by achieving the best results on specific datasets as reported in the Matbench study \cite{dunn2020benchmarking}, they often overfit the i.i.d. training datasets, which prevent them from achieving good performance on the OOD test sets. CGCNN's simplicity and primitiveness overcome this issue as it carries less bias from its design to perform well on OOD tests of some specific datasets (e.g., MaterialsProject formation energy/ band-gap dataset, etc.). This is why it outperformed all the SOTA GNN algorithms in the Matbench study on the OOD property prediction tasks for the dielectric dataset. However, with the dataset size increasing, its performance started to lag behind ALIGNN and DeeperGATGNN. ALIGNN's SOTA performance can be attributed to its line graph encoding used to incorporate the triplet feature and the two-step edge-gated convolution operation. On the other hand, DeeperGATGNN's unique architecture based on a global attention mechanism aided with differentiable group normalization and skip-connection contributes to its overall SOTA performance on the perovskites dataset. Despite being the best structure-based GNN models in the current MatBench leaderboard~\cite{matbench_leaderboard}, coGN and coNGN failed to outperform CGCNN, ALIGNN, and DeeperGATGNN on OOD prediction for any of the datasets. But they outperformed the rest of the algorithms as they are also designed to adapt well for some particular datasets. This indicates that OOD data techniques, such as domain adaptation, are needed to alleviate their prediction performance. Moreover, we found that MEGNet, SchNet, and DimeNet++ achieved worse but similar OOD performances on all three datasets, which demonstrates they are not suitable GNN models for making OOD materials property predictions.

\begin{figure}[!htb] 
    \centering
    \begin{minipage}[c]{0.6\textwidth}
        \centering
        \includegraphics[width=\textwidth]{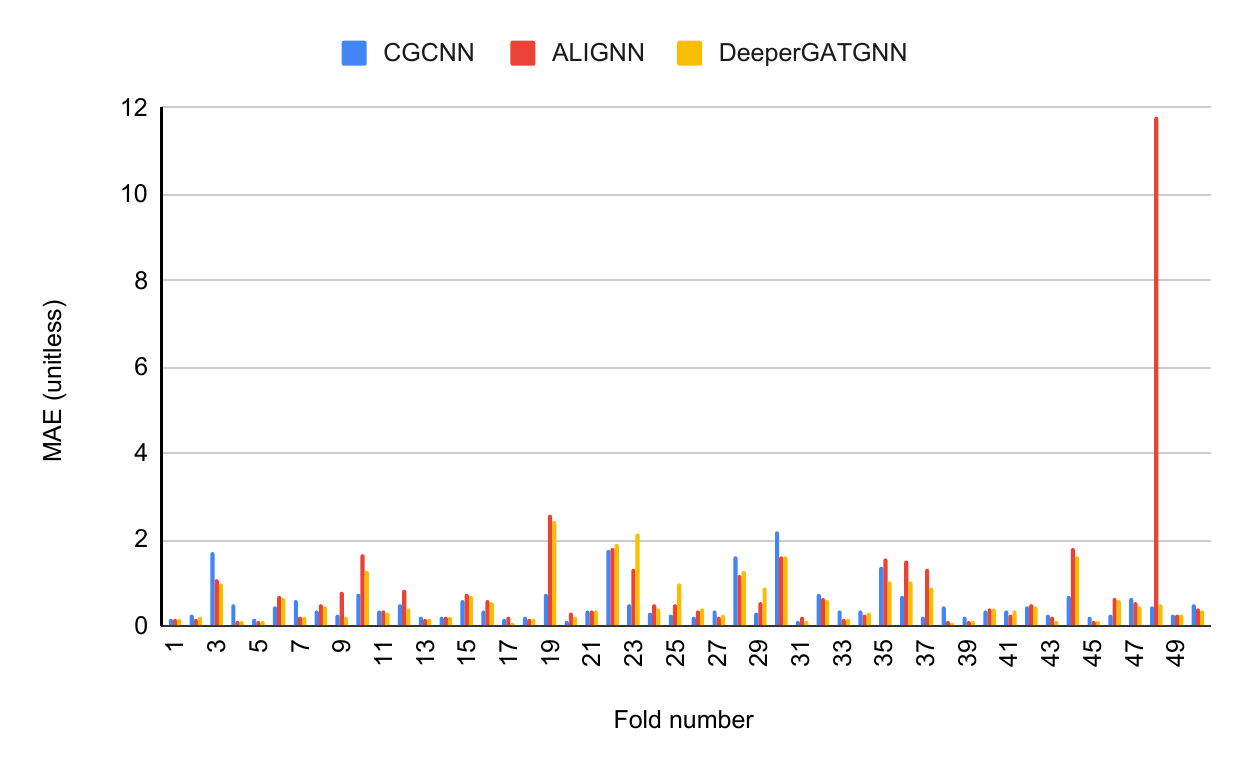}
        \subcaption{LOCO}
        \label{fig:di_foldwise_lo}
        \vspace{-1pt}       
    \end{minipage}\\
    \begin{minipage}[c]{0.495\textwidth}
        \centering
        \includegraphics[width=\textwidth]{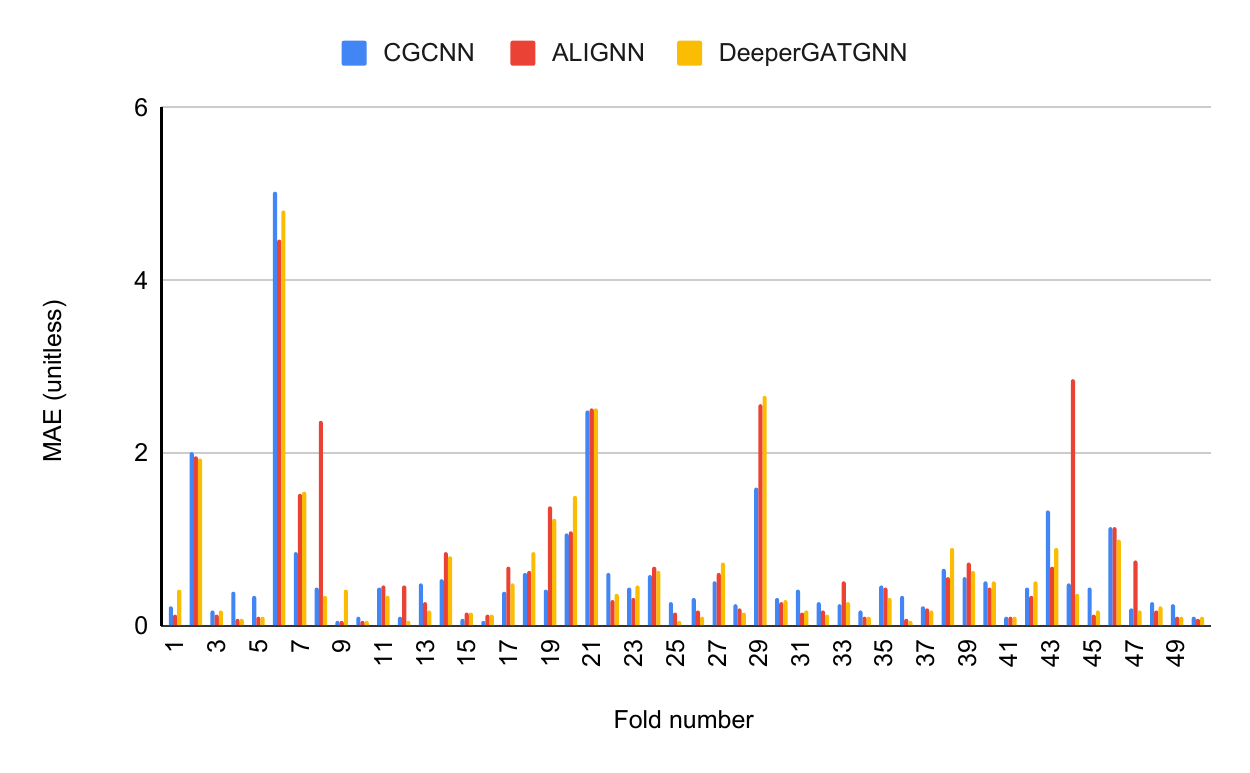}
        \subcaption{SparseXcluster}
        \label{fig:di_foldwise_xc}
        \vspace{-1pt}       
    \end{minipage}
    \begin{minipage}[c]{0.495\textwidth}
        \centering
        \includegraphics[width=\textwidth]{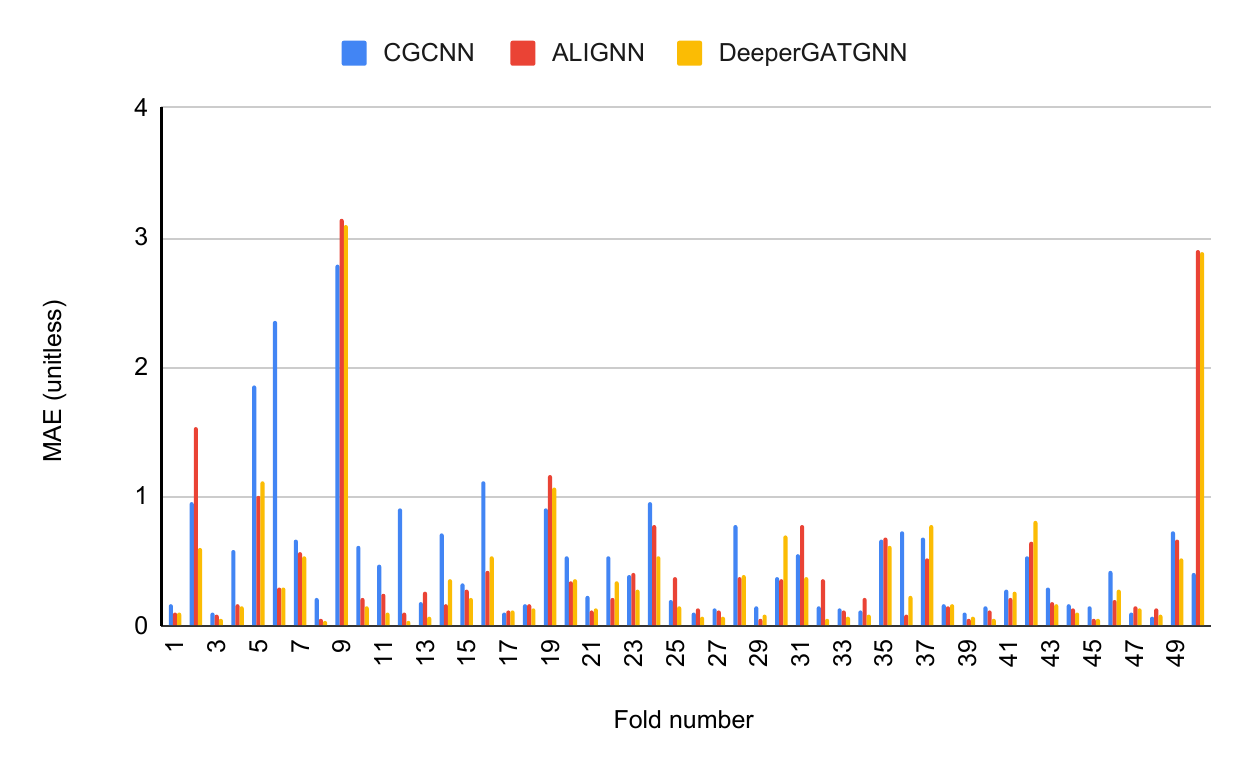}
        \subcaption{SparseYcluster}
        \label{fig:di_foldwise_yc}
        \vspace{-1pt}       
    \end{minipage}\\
    \begin{minipage}[c]{0.495\textwidth}
        \centering
        \includegraphics[width=\textwidth]{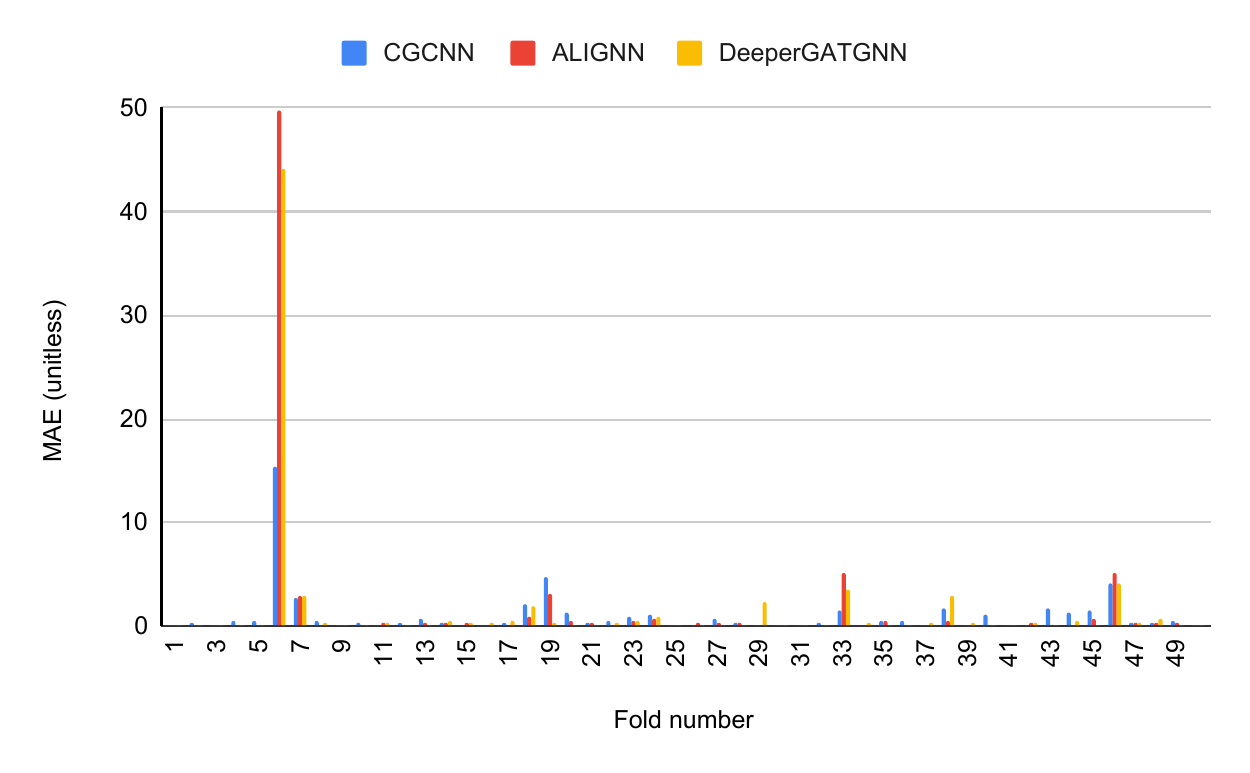}
        \subcaption{SparseXsingle}
        \label{fig:di_foldwise_xs}
        \vspace{-1pt}       
    \end{minipage}
    \begin{minipage}[c]{0.495\textwidth}
        \centering
        \includegraphics[width=\textwidth]{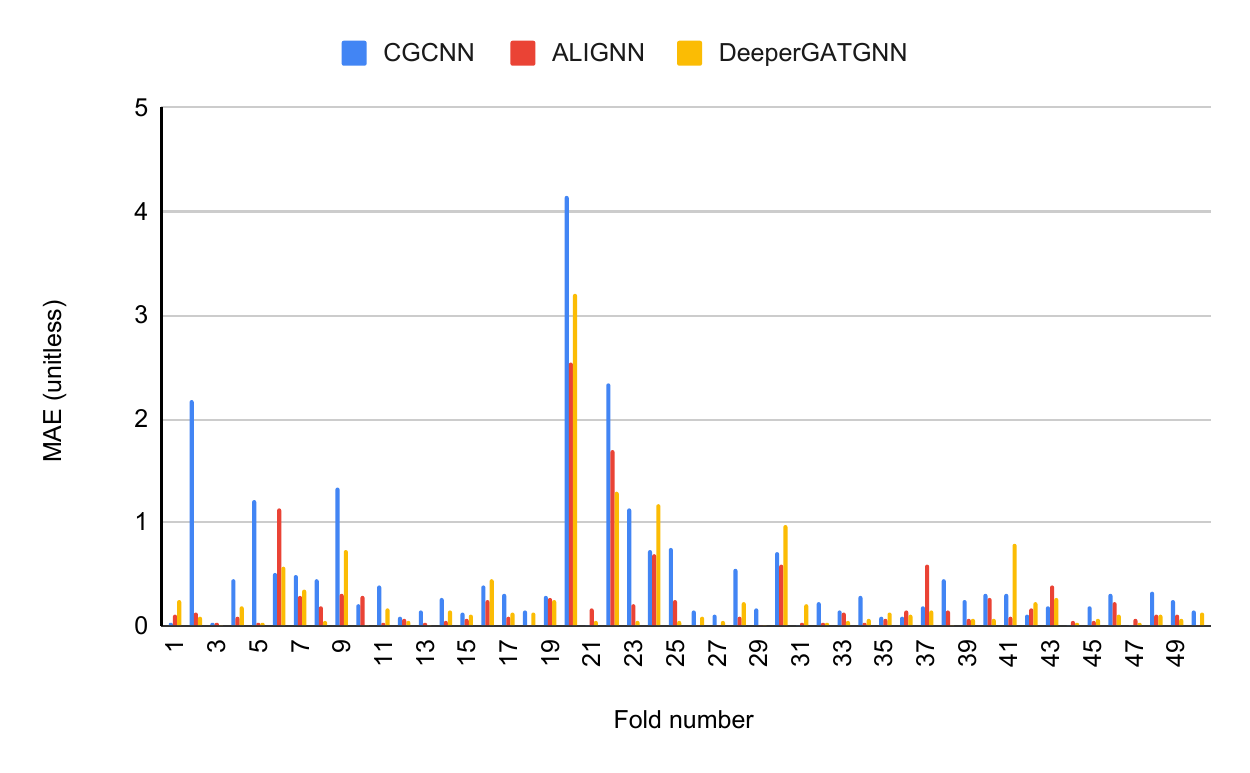}
        \subcaption{SparseYsingle}
        \label{fig:di_foldwise_ys}
        \vspace{-1pt}   
    \end{minipage}

     \caption{\textbf{Distribution of the MAEs for each fold of CGCNN, ALIGNN, and DeeperGATGNN on the dielectric dataset for (a) LOCO, (b) SparseXcluster, (c) SparseYcluster, (d) SparseXsingle, and (e) SparseYsingle OOD targets.}\\It showed that a few folds/clusters are extremely difficult to predict with MAE values greater than 1.0, which lead to high variation in the models' predictions.}
     \label{fig:di_foldwise}
\end{figure}

Although CGCNN, ALIGNN, and DeeperGATGNN displayed high resilience in handling OOD test data, their results are still bottlenecked by poor results for a few test clusters. The fold-wise MAE plots for these three algorithms on the dielectric dataset, elasticity dataset, and perovskites dataset are presented in Fig.~\ref{fig:di_foldwise}, Supplementary Fig. S3, and S4, respectively. The distribution of the MAE for 50 folds/clusters showed that only a few clusters are responsible for the overall MAE of each algorithm to surge. We also find that while ALIGNN achieves best or the second best OOD performance on the dielectric datasets (Table~\ref{table:dielectric}), it can have significantly degraded prediction MAEs for a few OOD test sets as shown in the highest peak in Fig.~\ref{fig:di_foldwise} (a - e). This analysis highlights specific areas where each algorithm's performance could be further optimized to enhance its overall accuracy and reliability. The parity plots of CGCNN, ALIGNN, and DeeperGATGNN's prediction performance on the perovskites dataset are plotted in Fig.~\ref{fig:parity_loco} for the LOCO targets, and in Supplementary Fig. S5 - S8 for the rest of the targets. These figures demonstrated superior OOD prediction performance of DeeperGATGNN compared to CGCNN and coGN on the perovskites dataset. For all categories of targets, DeeperGATGNN achieved significantly better prediction accuracy compared to CGCNN and coGN for non-OOD samples, which is proportional to their prediction accuracy for the OOD samples.

\begin{table}[!htb]
\caption{50-fold cross-validation MAEs (unitless) of different GNN models on the dielectric dataset for five different types of OOD problems. The best results, second best results, and worst results are marked by bold letters, underlines, and parentheses, respectively.}
\begin{center}
\begin{tabular}{l c c c c c}
\hline
\rowcolor[HTML]{C0C0C0}
\textbf{Algorithm} & 
\textbf{LOCO} & 
\textbf{SparseXcluster} &
\textbf{SparseXsingle} & 
\textbf{SparseYcluster} &
\textbf{SparseYsingle} \\ \hline

CGCNN &	\textbf{0.5144} &	\textbf{0.6006} &	\textbf{0.9888}	 & 0.5254	 & 0.4777 \\
MEGNet &	2.6830  &	2.8044  &	3.8384  &	2.5918  &	2.5706 \\
SchNet &	2.7074 &	2.8368 &	(3.9767) &	(2.5988) &	2.5566 \\
DimeNet++ &	(2.7720) &	(2.9378) &	3.8629 &	2.5947 &	(2.5866) \\
ALIGNN &	0.8592 &	\underline{0.6953} &	\underline{1.5115} &	\underline{0.4359} &	\textbf{0.2513} \\
DeeperGATGNN &	\underline{0.5911} &	1.4056 &	1.5755 &	\textbf{0.3959} &	\underline{0.2733} \\
coGN &	1.3286 &	1.3277 &	2.3958 &	1.0878 &	0.9051 \\
coNGN &	1.3365 &	1.3279 &	2.3956 &	1.0878 &	0.9054 \\
\hline

\end{tabular}
\label{table:dielectric}
\end{center}
\end{table}

\begin{table}[!htb]
\caption{50-fold cross-validation MAEs ($\log_{10}$(GPa)) of different GNN models on the elasticity dataset for five different types of OOD problems. The best results, second best results, and worst results are marked by bold letters, underlines, and parentheses, respectively.}
\begin{center}
\begin{tabular}{l c c c c c}
\hline
\rowcolor[HTML]{C0C0C0}
\textbf{Algorithm} & 
\textbf{LOCO} & 
\textbf{SparseXcluster} &
\textbf{SparseXsingle} & 
\textbf{SparseYcluster} &
\textbf{SparseYsingle} \\ \hline

CGCNN &	\textbf{0.0585} &	\textbf{0.0499} &	\underline{0.0895} &	\underline{0.0752} &	0.0840 \\
MEGNet &	(1.4468) &	(1.4113) &	1.3099 &	(1.5659) &	1.4491 \\
SchNet &	1.4065 &	1.3455 &	1.2363 &	1.5592 &	(1.4855) \\
DimeNet++ &	1.4242 &	1.3562 &	(1.3214) &	1.5454 &	1.4828 \\
ALIGNN &	\underline{0.0974} &	\underline{0.0834} &	\textbf{0.0853} &	\textbf{0.0631} &	\textbf{0.0450} \\
DeeperGATGNN &	0.1173 &	0.1109 &	0.1140 &	0.0858 &	\underline{0.0807} \\
coGN &	0.4306 &	0.4207 &	0.2831 &	0.3996 &	0.3188 \\
coNGN &	0.4309 &	0.4207 &	0.2833 &	0.3997 &	0.3186 \\
\hline

\end{tabular}
\label{table:elasticity}
\end{center}
\end{table}

\begin{table}[!htb]
\caption{50-fold cross-validation MAEs (eV/unit cell) of different GNN models on the perovskites dataset for five different types of OOD problems. The best results, second best results, and worst results are marked by bold letters, underlines, and parentheses, respectively.}
\begin{center}
\begin{tabular}{l c c c c c}
\hline
\rowcolor[HTML]{C0C0C0}
\textbf{Algorithm} & 
\textbf{LOCO} & 
\textbf{SparseXcluster} &
\textbf{SparseXsingle} & 
\textbf{SparseYcluster} &
\textbf{SparseYsingle} \\ \hline

CGCNN &	0.0651 &	0.0839 &	0.0689 &	0.0740 &	0.0768 \\
MEGNet &	1.4654 &	1.4485 &	1.5005 &	1.4685 &	1.5048 \\
SchNet &	1.4644 &	1.4509 &	1.5133 &	(1.4736) &	(1.5173) \\
DimeNet++ &	 (1.4666) &	(1.4567) &	(1.5248) &	1.4732 &	1.5148 \\
ALIGNN & \underline{0.0386} & \underline{0.0468} &	\underline{0.0457} & \underline{0.0341} &	\textbf{0.0243} \\
DeeperGATGNN & \textbf{0.0365} &	\textbf{0.0464} & \textbf{0.0373} & \textbf{0.0333} &	\underline{0.0259} \\ 
coGN &	0.8099 &	0.8598 &	0.9436 &	0.7738 &	0.9561 \\
coNGN &	0.8092 &	0.8598 &	0.9435 &	0.7734 &	0.9560 \\
\hline

\end{tabular}
\label{table:perovskites}
\end{center}
\end{table}

\begin{figure}[!htb]
    \centering
    \includegraphics[width=\textwidth]{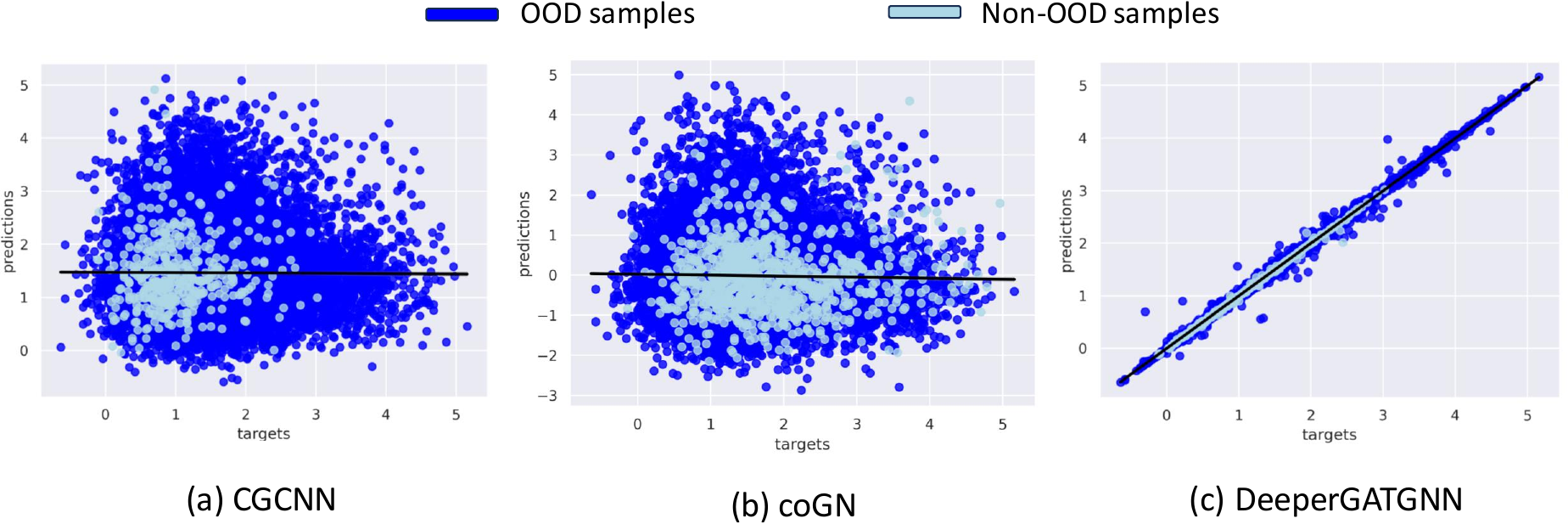}
    \caption{\textbf{Parity plots for both OOD and non-OOD samples for the LOCO target generation method on the perovskites dataset.}\\These show that DeeperGATGNN has better performance on the non-OOD samples compared to CGCNN and coGN, which is proportional to its better OOD performance than the other two for the LOCO targets.}
    \label{fig:parity_loco}
\end{figure}

\subsection{Performance comparison with MatBench SOTA performance}
\begin{figure}[!htb] 
    \centering
    \begin{minipage}[c]{0.6\textwidth}
        \centering
        \includegraphics[width=\textwidth]{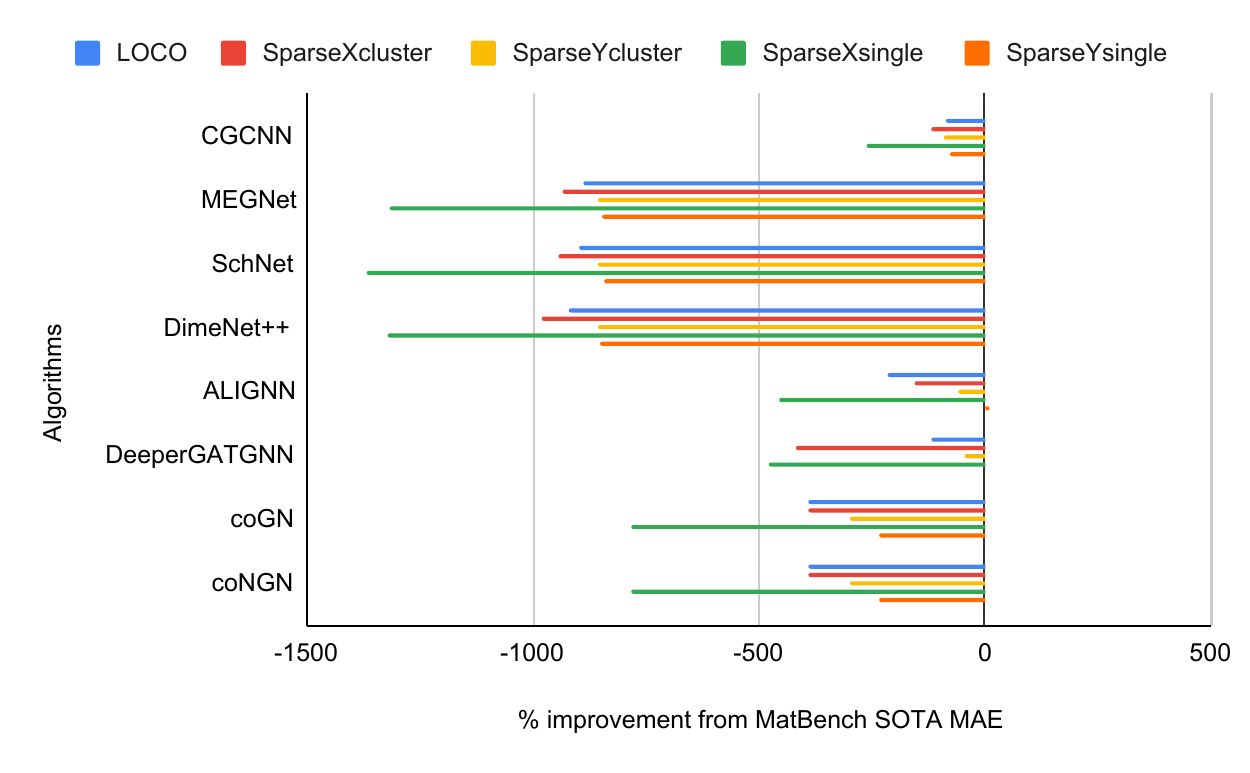}
        \subcaption{Dielectric dataset.}
        \label{fig:di_sota}
        \vspace{-1pt}       
    \end{minipage}\\
    \begin{minipage}[c]{0.495\textwidth}
        \centering
        \includegraphics[width=\textwidth]{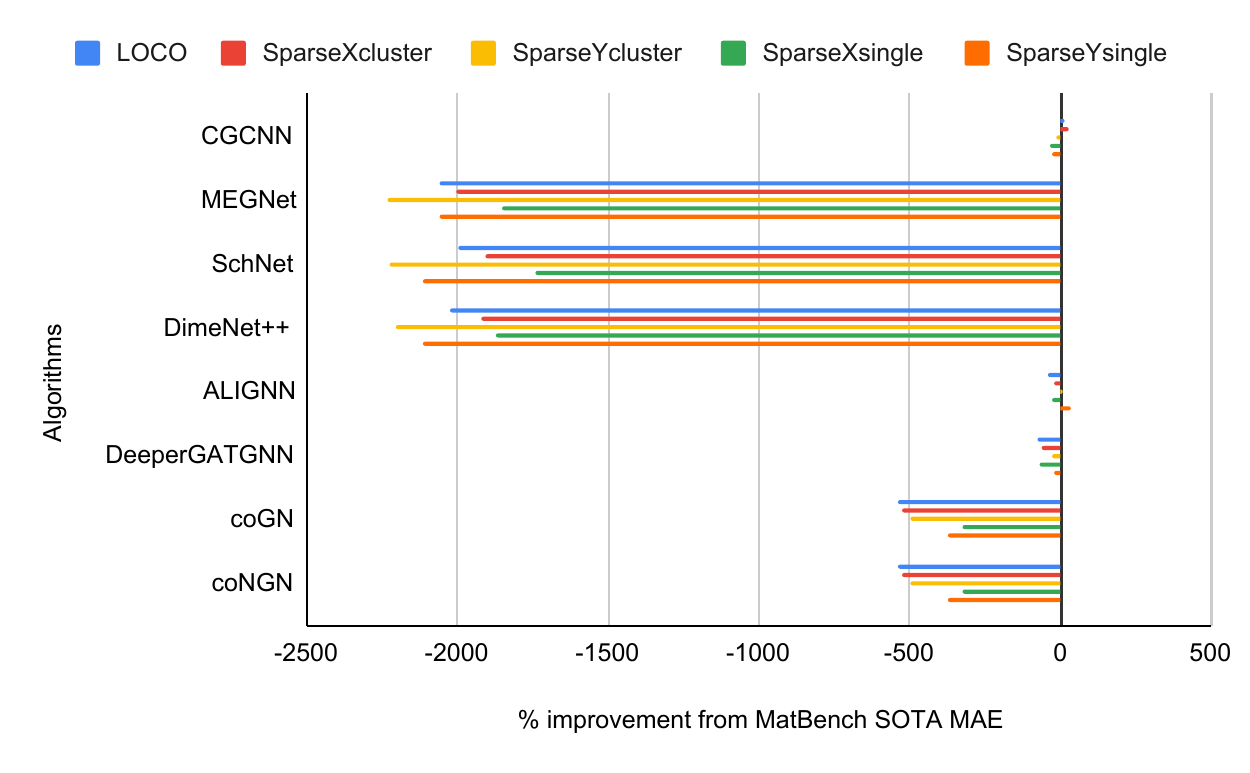}
        \subcaption{Elasticity dataset.}
        \label{fig:el_sota}
        \vspace{-1pt}       
    \end{minipage}
    \begin{minipage}[c]{0.495\textwidth}
        \centering
        \includegraphics[width=\textwidth]{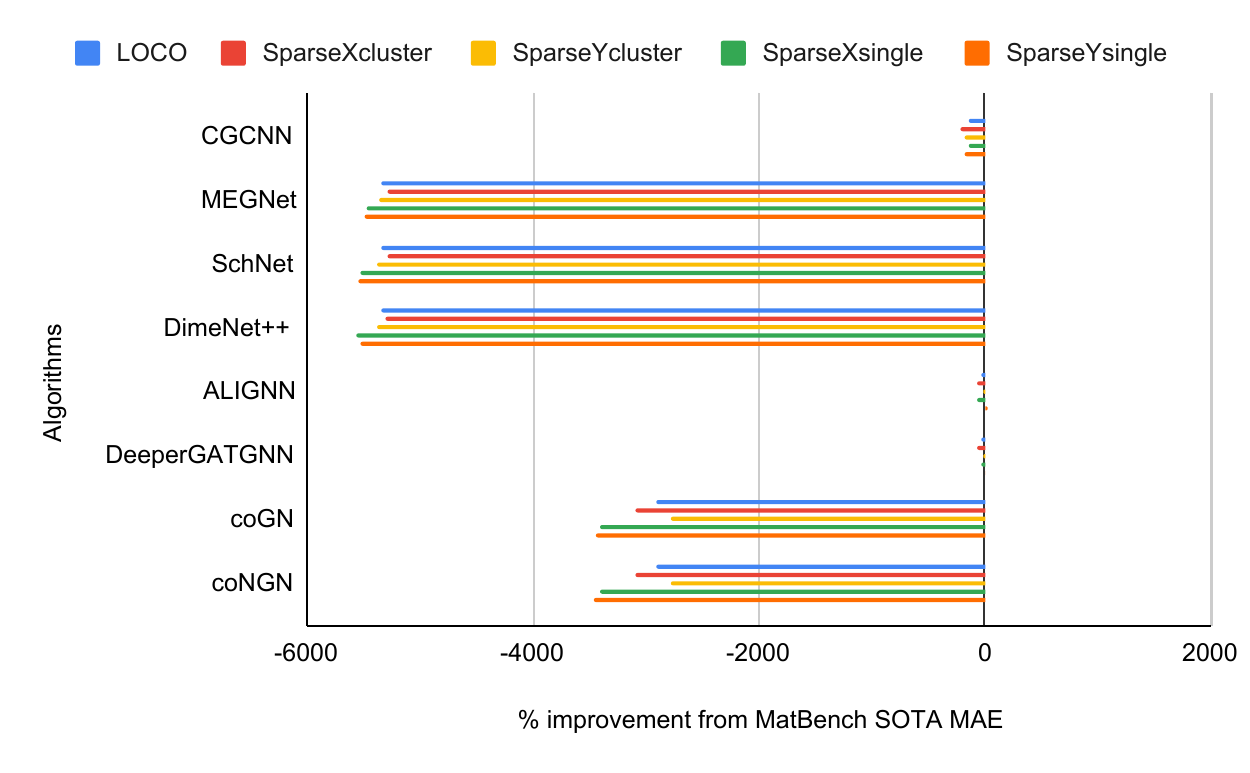}
        \subcaption{Perovskites dataset.}
        \label{fig:pe_sota}
        \vspace{-1pt}       
    \end{minipage}

     \caption{\textbf{Performance comparison of different GNN models' MAEs for all five different types of OOD targets with the SOTA MAEs found in the MatBench study on the (a) dielectric dataset, (b) elasticity dataset, and (c) perovskites dataset.}\\Although on average, all models' MAEs are significantly higher than the SOTA MAEs, CGCNN, ALIGNN, and DeeperGATGNN outperformed MatBench's SOTA results in some cases.}
     \label{fig:comp_sota}
\end{figure}

To investigate the issue of performance degradation of GNN models in OOD material property prediction, we compare the MatBench SOTA algorithms' performance on the OOD datasets with those on the Matbench study (Fig.~\ref{fig:di_sota}). 
The MatBench SOTA algorithms and their MAEs for all three datasets in the Matbench study \cite{dunn2020benchmarking} are listed in Table~\ref{tab:datasets}. 
We calculated the performance changes (in percentage) from the MatBench SOTA MAE to the MAE found for all five OOD target generation methods for each algorithm. The goal of this analysis is to find out the feasibility of current GNN algorithms for high-performance OOD materials property prediction. We found that all models' OOD performances are significantly worse than their SOTA results in MatBench, with degradation ranging from -0.83\% to a substantial -1366.87\%. These results indicate the inadequacy of current GNN algorithms for OOD property prediction for materials data. The only exception found on the dielectric dataset is for the SparseYsingle targets, where ALIGNN's performance is found to be improved by 7.31\%. On the other hand, MEGNet, SchNet, and DimeNet++ performed the worst with a performance change of > -750\% for all five types of OOD targets. However, judging from Fig.~\ref{fig:di_sota}, we observed that CGCNN adapted the best on average in the new OOD target-based predictions on the dielectric dataset.

Similar comparison results on the elasticity dataset are shown in Fig.~\ref{fig:el_sota}. Again, all the GNN algorithms exhibited their deficiency to generalize, achieving the MAE increases for all algorithms across all OOD test sets ranging from -12.24\% to a remarkable -2237.23\%. Exception from these results are CGCNN's outperforming the MAEs of SOTA algorithms for LOCO and SparseXcluster targets (12.7\%, and 25.51\% improvement, respectively), and ALIGNN's superior performance for both the SparseYcluster and SparseYsingle targets (5.82\%, and 32.82\% improvement, respectively). The finding from this figure is that with the increase of data from the dielectric dataset to the elasticity dataset, the previously deteriorated results of MEGNet, SchNet, and DimeNet++ became even more deteriorated with a notable performance degradation of > -1500\% for all five types of OOD targets. However, DeeperGATGNN, CGCNN, and ALIGNN's average performance degradations on the elasticity dataset were lower than those of the dielectric dataset.

Finally, we showed the performance degradation on the perovskites dataset of the MAEs for all MatBench SOTA algorithms by comparing the OOD results in Fig.~\ref{fig:pe_sota} and their i.i.d. performance in Table \ref{tab:datasets}. We again noticed that almost all algorithms' performance is much worse than their performance on the MatBench study, ranging from -23.84\% to a staggering -5568.31\%. In contrast, ALIGNN and DeeperGATGNN outperformed the previous SOTA MAE by 3.87\%, and 9.73\%, respectively, for the SparseYsingle targets. We observed that ALIGNN improved on all three datasets for the SparseYsingle OOD targets, which demonstrated its fair generalization capability for this type of OOD target. But considering all the OOD target generation methods, none of the algorithms showed good generalization capability on average, demonstrating the necessity for methods like domain adaptation to improve the OOD prediction performance of current GNN models.

\subsection{Comparison of OOD performance with baseline i.i.d. performance}
\begin{figure}[!htb] 
    \centering
    \begin{minipage}[c]{0.6\textwidth}
        \centering
        \includegraphics[width=\textwidth]{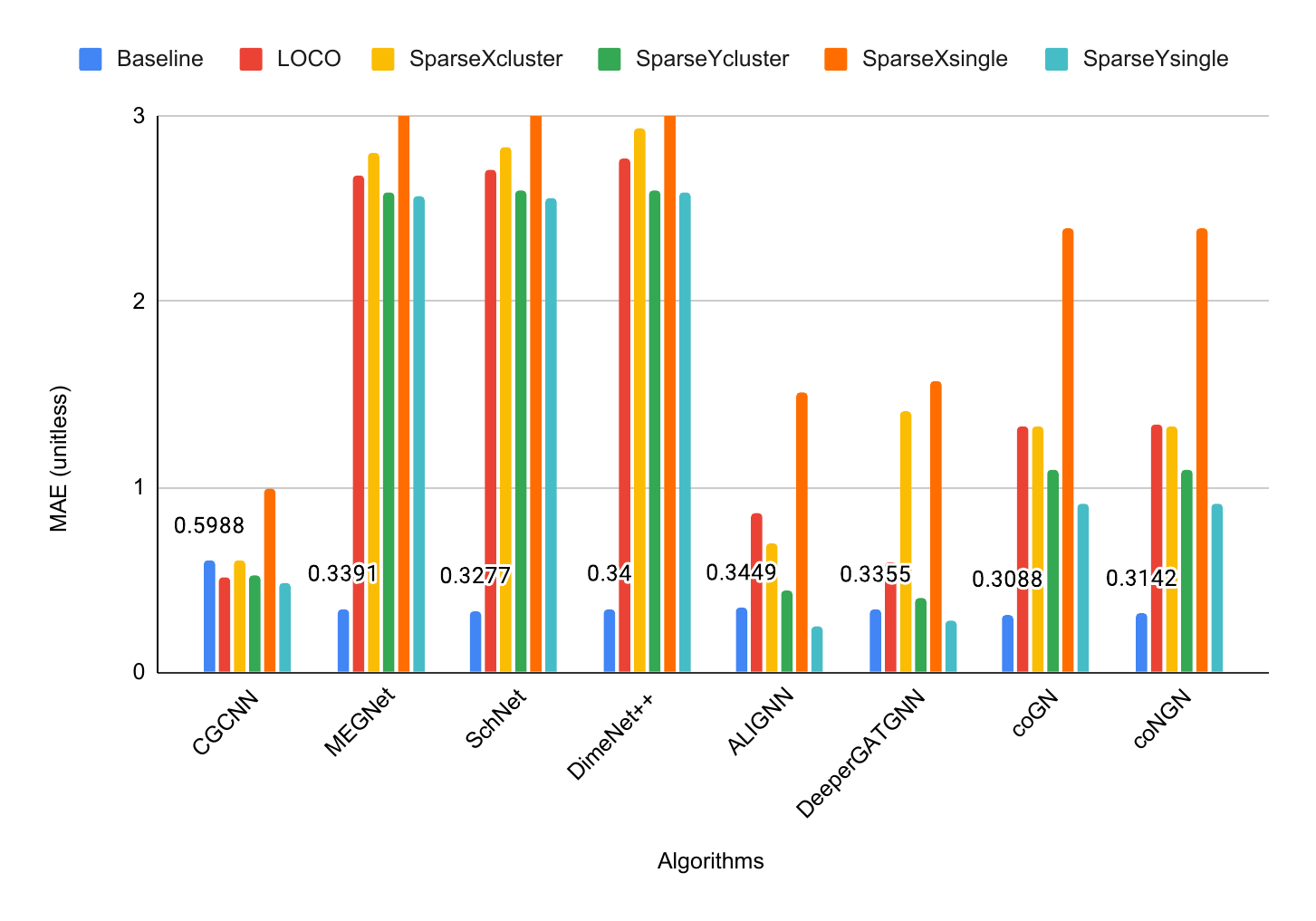}
        \subcaption{Dielectric dataset.}
        \label{fig:di_baseline}
        \vspace{-1pt}       
    \end{minipage}\\
    \begin{minipage}[c]{0.495\textwidth}
        \centering
        \includegraphics[width=\textwidth]{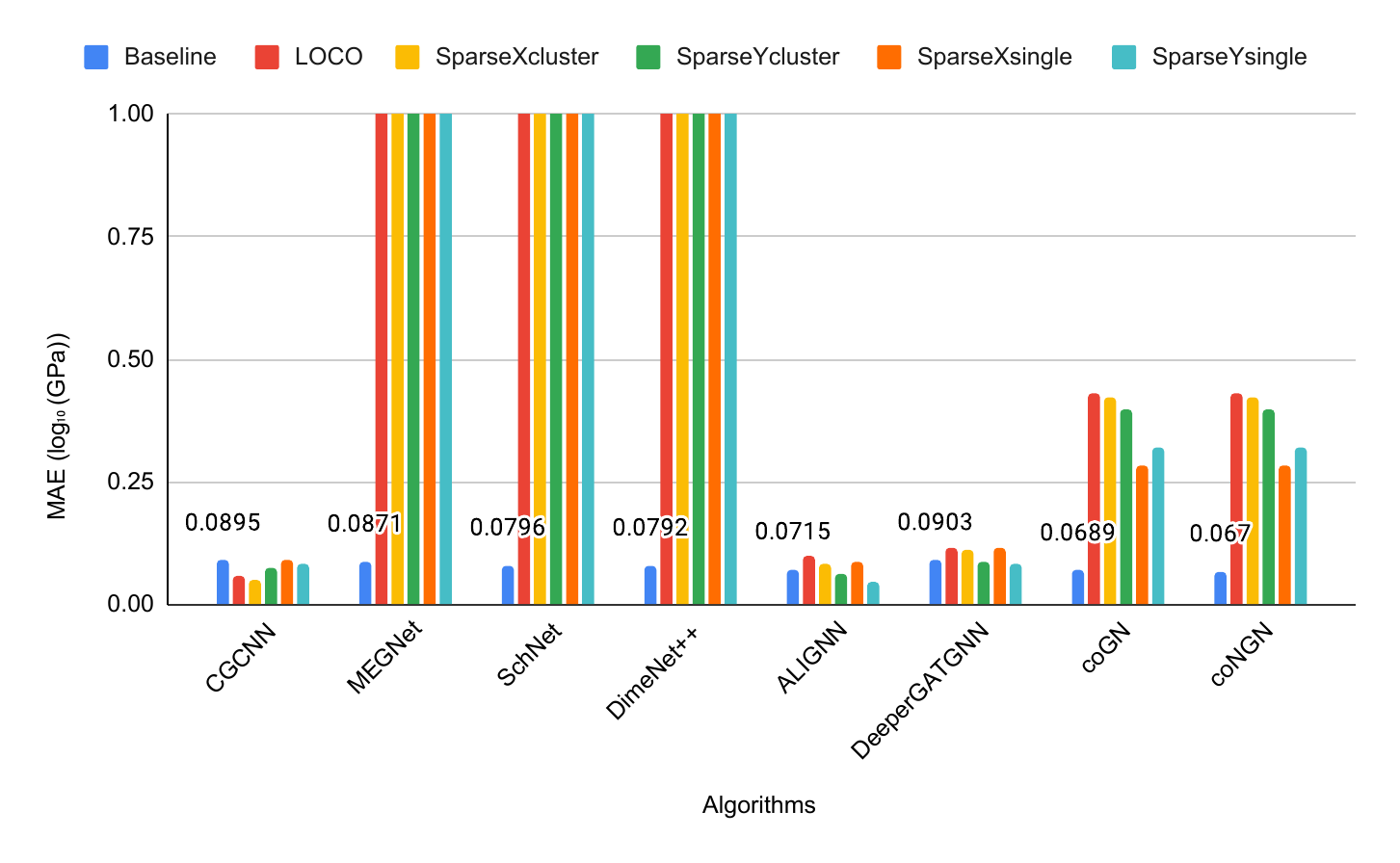}
        \subcaption{Elasticity dataset.}
        \label{fig:el_baseline}
        \vspace{-1pt}       
    \end{minipage}
    \begin{minipage}[c]{0.495\textwidth}
        \centering
        \includegraphics[width=\textwidth]{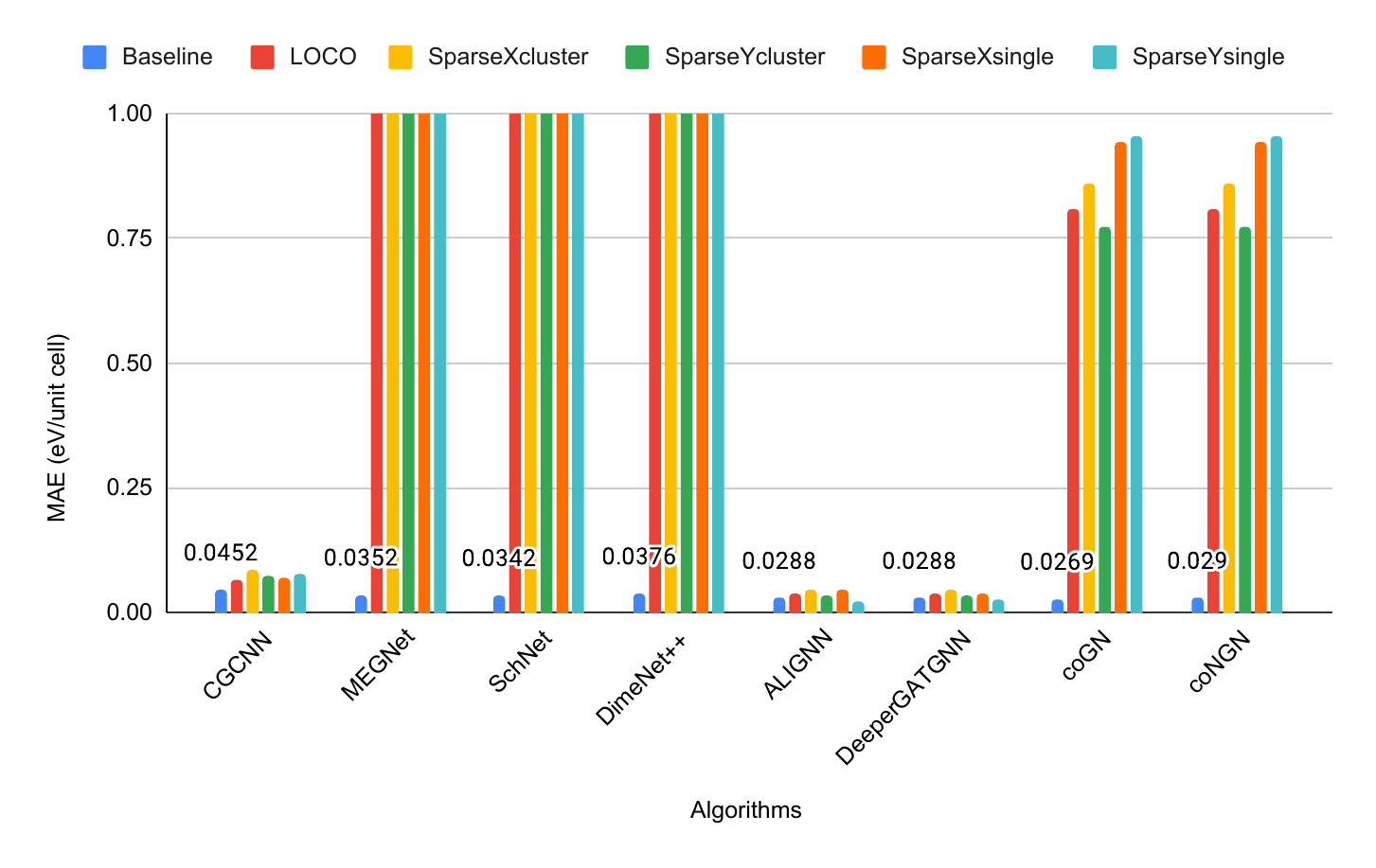}
        \subcaption{Perovskites dataset.}
        \label{fig:pe_baseline}
        \vspace{-1pt}       
    \end{minipage}

     \caption{\textbf{Performance comparison of different GNN models' MAEs for all five different types of OOD targets with their baseline i.i.d. MAEs found in the MatBench study on the (a) dielectric dataset, (b) elasticity dataset, and (c) perovskites dataset.}The baseline MAE for each algorithm is labeled. All the models on average achieved higher MAEs than their baseline i.i.d. MAEs which proved the inadequacy of current GNN models for OOD materials property prediction.}
     \label{fig:comp_baseline}
\end{figure}

Here we aim to check how the evaluated GNNs' performances degrade when changing their test sets from i.i.d to OOD. 
The i.i.d. baseline MAEs for all GNN algorithms except DeeperGATGNN can be found on the MatBench leaderboard~\cite{matbench_leaderboard} while the DeeperGATGNN's baseline result was obtained by our experiments using the same test set splitting as Matbench. The comparison of all algorithms' prediction performances for all five OOD targets with their i.i.d. baseline MAEs in the MatBench study on the dielectric dataset are shown in Fig.~\ref{fig:di_baseline}. We limited the maximum value of $y$-axis to $3$ for better visualization. We found that OOD MAEs of MEGNet, SchNet, DimeNet++, coGN, and coNGN are significantly worse than their i.i.d. baseline MAEs for all OOD target sets, which proved their inadequate prediction capabilities on OOD datasets. This was evident from the results of Subsection~\ref{subsec:comp_ood} as all of the SOTA OOD results were achieved by either CGCNN, ALIGNN, or DeeperGATGNN. However, CGCNN was found to be most benefited from the 50-fold OOD test set cross validation experiments as it improved over its baseline performance in the MatBench study on three targets - LOCO (14.09\%), SparseYcluster (12.26\%), and SparseYsingle (20.23\%), while ALIGNN, and DeeperGATGNN only managed to improve for the SparseYsingle targets (27.14\%, and 18.53\%, respectively). 

The OOD test set comparison with the MatBench baseline results of each algorithm on the elasticity and perovskites dataset are shown in Fig.~\ref{fig:el_baseline}, and \ref{fig:pe_baseline}, respectively. We again limited the maximum value of $y$-axis to $1$ for both these figures for better visualization. On the elasticity dataset, CGCNN improved over its baseline MAE for all OOD target sets (LOCO: 34.65\%, SparseXcluster: 44.24\%, SparseYcluster: 15.98\%, SparseXsingle: 0.01\%, SparseYsingle: 6.18\%), while it failed to improve over its baseline performance on the perovskites dataset for any type of OOD targets. ALIGNN, and DeeperGATGNN's OOD results are better than their baseline results for only the SparseYcluster targets (11.75\%, and 5\%, respectively), and SparseYsingle targets (37.05\%, and 10.65\%, respectively) on the elasticity dataset. However, they only succeeded in improving over their baseline MAEs for the SparseXsingle targets (15.68\%, and 10.21\%, respectively). In contrast, all other models performed significantly worse than these three on both datasets.

Our 50-fold cross validation scenario is a much more complex challenge for the current GNN algorithms than the scenario presented in the MatBench Study because of the OOD test sets for each fold rather than the i.i.d. test sets from random splitting as used in Matbench. Despite this, three algorithms' success in outperforming their baseline results for multiple targets proved the robustness of their inherent prediction capabilities. It also manifested the lack of reliability of the material property prediction performance of current GNNs reported in the MatBench study as indicators of their effectiveness in real-world materials property prediction, especially for those obscure materials that researchers desire.

\subsection{Physical insights}
\begin{figure}[!htb]
    \centering
    \includegraphics[width=\textwidth]{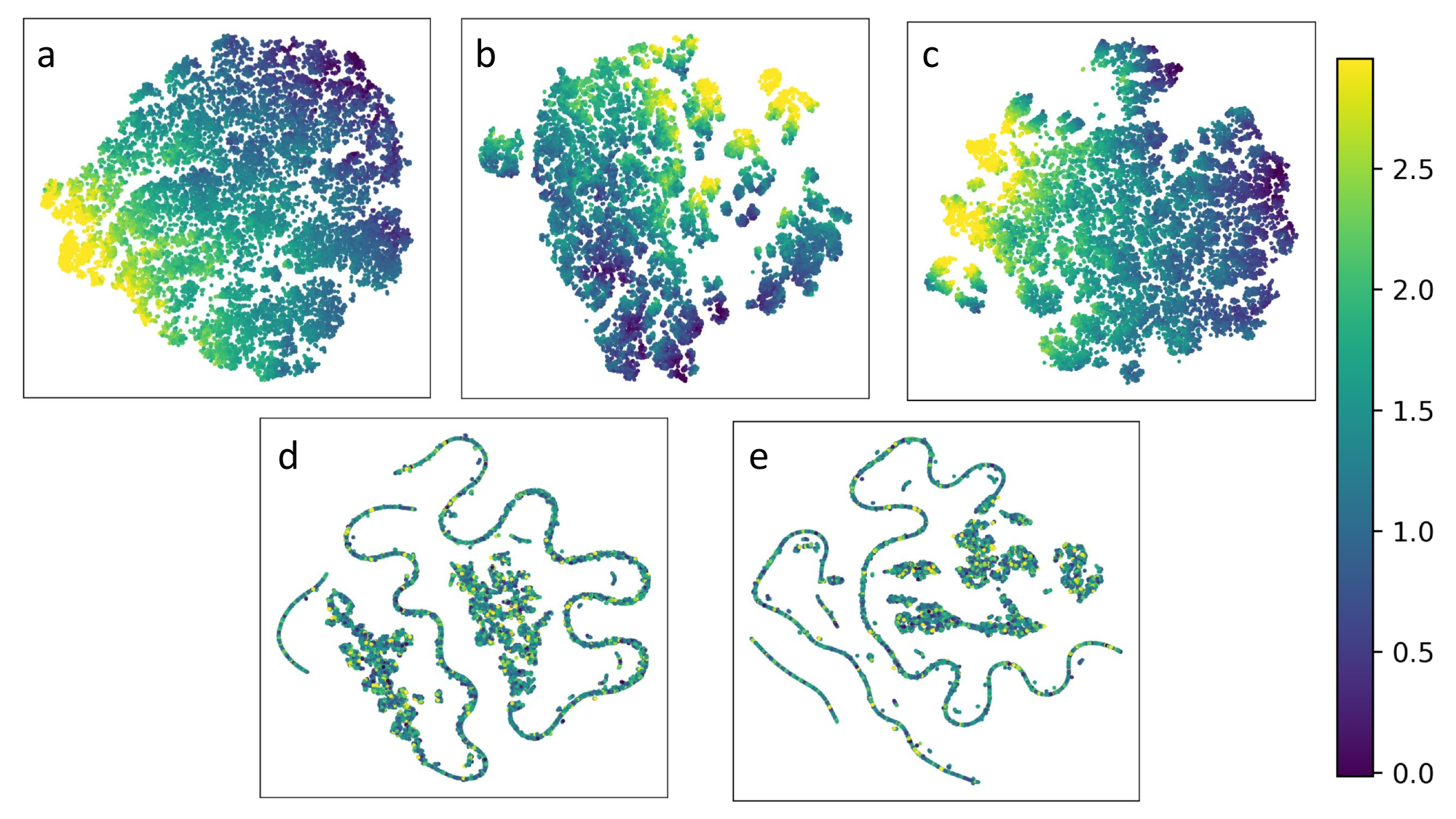}
    \caption{\textbf{Formation energy distribution of perovskites materials in the latent space.}\\t-SNE diagrams for (a) CGCNN, (b) ALIGNN, (c) DeeperGATGNN, (d) coGN, and (e) coNGN were plotted after training with the perovskites dataset and retrieving latent representation of the first layer after the final graph convolution layer. Different colors indicate different formation energy levels in the latent space, where each point represents a separate perovskite material. CGCNN, ALIGNN, and DeeperGATGNN have much smoother clusters compared to coGN and coNGN, where different colored clusters are hardly separable.}
    \label{fig:tsne}
\end{figure}

We utilized t-distributed stochastic neighbor embedding (t-SNE)~\cite{van2008visualizing}, a commonly utilized non-linear method for visualizing and interpreting complex, high-dimensional data, to investigate specific insights into the materials' physics. t-SNE aims to reduce higher-dimensional data into a much lower dimension (typically 2D or 3D) while preserving the proximity of data points in both dimensions. We selected the perovskites formation energy dataset for training. We also only investigate CGCNN, ALIGNN, DeeperGATGNN, coGN, and coNGN for this experiment - the first three models for their comparatively robust OOD prediction performance among all GNN models, and the last two for their SOTA performance in the current MatBench leaderboard~\cite{matbench_leaderboard}. Our objective is to visualize the distribution of latent representations learned through the training of different models. For each trained model, we retrieved the output of the first layer after the final graph convolution layer and plotted the t-SNE diagrams in Fig.~\ref{fig:tsne}.

The t-SNE diagrams portray integrated latent spaces that combine structure and composition information for the materials that have been trained. Various colors correspond to different levels of formation energy for the samples represented in those latent spaces. We noticed that each of these GNNs is capable of producing effective representations that result in the clustering of materials with similar formation energies. Within each cluster, it can be anticipated that points will exhibit resemblances in both their atomic configurations and elemental compositions. While each model may produce distinct latent spaces, we can gain valuable insights into their prediction patterns by examining these clusters. For example, we observed that coGN and coNGN have similar patterns in their latent spaces which can explain their similar OOD benchmark results (see Table~\ref{table:perovskites}). Also, we can see that the lower and higher energy distribution overlaps in the latent spaces of coGN and coNGN are very high. We can hardly separate different colored regions in their latent spaces, whereas the same separation is much smoother in CGCNN, ALIGNN, and DeeperGATGNN. This might explain the SOTA OOD performance of CGCNN, ALIGNN, and DeeperGATGNN, and the poor OOD performance of coGN and coNGN.

\FloatBarrier

\section{Discussion}

Our benchmark study aimed to systematically evaluate the performance of eight GNN models in the challenging task of structure-based OOD materials property prediction and identify models that exhibit superior OOD prediction performance while understanding the factors contributing to their efficacy. The motivation behind this work stems from the fact that material researchers are typically drawn to exploring novel materials with exceptional properties and unconventional compositions or structures, which presents a common difficulty in current ML techniques known as the OOD prediction problem. Through rigorous experimentation, we found that no single algorithm achieved SOTA performance for all OOD target set generation methods for a given dataset, let alone on all datasets. The reason is severalfold.

First, to ensure a fair and comprehensive evaluation, we subjected each of the three chosen datasets from the MatBench study to a systematic division into 50 folds. This division process was conducted using five distinct target set generation methods, resulting in a total of 250 folds across the three datasets. Notably, within each fold, the test samples were methodically drawn to be OOD from the remaining dataset using a strategic approach based on predefined criteria. These posed a strong challenge for the GNNs chosen in this study as they were not only tasked with performing effectively on the challenging OOD test samples but also with demonstrating effectiveness across the entirety of the 50-fold cross-validation setup, underscoring the robustness and adaptability required for this comprehensive assessment.

Second, we found that no single algorithm triumphed in all situations which indicated their lack of generalization capability across different datasets and unreliability in making real-world materials property prediction. However, CGCNN, ALIGNN, and DeeperGATGNN proved to be more robust than other algorithms. CGCNN excelled in certain OOD scenarios due to its rudimentary nature. This achievement is particularly significant given that CGCNN outperformed both coGN and coNGN, which currently hold the SOTA performance for the majority of tasks in the MatBench leaderboard. While cutting-edge GNNs often strive for optimal results on specific datasets, they risk getting prone to overfitting to these specific datasets or types of material data. CGCNN's simplistic and primitive architecture overcomes this problem by offering a significantly less complex model compared to other SOTA GNNs, such as MEGNet. This makes CGCNN more challenging to overfit, as the likelihood of overfitting tends to increase with the number of parameters in a deep neural network model. But in some cases, having a larger model with more parameters might be necessary to capture intricate patterns in complex material data, which becomes a trade-off for CGCNN in performance on both the MatBench study and our OOD study. As a result, CGCNN demonstrated robust prediction performance on the dielectric dataset on average but trailed ALIGNN, and DeeperGATGNN (which have more training parameters and better architectures) with the increasing number of materials in other datasets. The key to ALIGNN's remarkable performance lies in its distinctive line graph encoding strategy, which enables the utilization of triplet features that effectively capture long-range interactions between atoms. Moreover, the incorporation of two levels of edge-gated convolutions in updating both node and edge features also plays a pivotal role in its SOTA performance on the elasticity dataset on average. However, DeeperGATGNN claimed the SOTA performance on average on the largest dataset (perovskites), as both ALIGNN and CGCNN suffered from over-smoothing (a phenomenon that makes GNN node features almost similar after a certain graph convolution layer). DeeperGATGNN evades this issue by incorporating differentiable group normalization and residual skip-connections, which allows it to use > 50 graph convolution layers to extract deeper-level features from the encoded materials graphs. Despite all this, no single algorithm dominates for all types of OOD targets, which necessitates new algorithms such as domain adaptation, or meta-learning to improve the ML OOD prediction performance.

The best GNN models in the MatBench study are coGN and coNGN which performed significantly worse than ALIGNN, DeeperGATGNN, and CGCNN, but performed better than the rest of the GNNs (MEGNet, DimeNet++, and SchNet). This demonstrated that their SOTA performances in the MatBench leaderboard seem to be largely due to the overfitting the specific datasets' folds. It is peculiar that despite leveraging line graphs to utilize angle information, the results of coGN and coNGN are not as competitive as ALIGNN for OOD prediction. Also, the effect of nested line graph (coNGN) is almost non-existent as its performance advantage over coGN ranges from 0.0003\% to 0.5953\% only among datasets for different targets, which raises the question about the impact of nesting with the non-nested version (coGN) for OOD targets. SchNet and DimeNet++ are primarily designed for molecular property prediction which can be accounted for their subpar performance on both the MatBench leaderboard and our OOD prediction benchmark. Moreover, MEGNet has the most number of parameters among all the GNNs selected, making it the most prone to overfitting.

Through a thorough examination, we also observed a noteworthy trend across all datasets and target generation methods: the performance of each algorithm for such OOD test sets is consistently lower than those of the baselines established in the MatBench study, except for a few cases (see Fig.~\ref{fig:comp_baseline}). This collective under-performance demonstrated that traditional GNN models are not robust enough to handle OOD property prediction yet. This empirical evidence necessitates incorporating enhanced robustness methods in these algorithms, such as domain adaptation, or federated learning. Surprisingly, both ALIGNN and DeeperGATGNN performed better than their baseline results for the SparseYsingle OOD test sets on all datasets. In fact, other GNN algorithms also had the least performance degradation in making predictions for this method compared to the other four. In contrast, the SparseXsingle targets caused all the algorithms the highest performance degradation on average. Moreover, Fig.~\ref{fig:di_foldwise}, and Supplementary Fig. S2 and S3 showed that only a few clusters/samples are extremely difficult to predict, which leads to the high variation of the model's prediction performance. As these partitions are done based on the structure $x$, or property $y$ values of the t-SNE of the OFM feature space, this can be a good research direction to find out the opposite physical relation of both directions values to design a more robust GNN. Of course, the main research goal of this work is to design highly robust GNN algorithms that achieve high-performance predictions on unknown outlier materials. The unexpected resilience displayed by CGCNN, ALIGNN, and DeeperGATGNN in a few cases can be a promising research direction to investigate further for this endeavor.

\section{Conclusion}
Due to material scientists' aspiration for novel exceptional materials, we conducted the first benchmark work that empirically investigated the feasibility of current graph neural network (GNN) algorithms for predicting properties of out-of-distribution (OOD) materials (materials that deviate from the distribution of the training set), which complements the related work for organic materials \cite{shimakawa2024extrapolative}. We formulated five categories of OOD problems using three inorganic material datasets from the MatBench study. Our rigorous experiments revealed significant generalization gaps in current state-of-the-art (SOTA) GNN algorithms, demonstrating their underperformances on OOD tasks compared to their baseline performances in the MatBench study. We showed that their underperformances primarily stem from the challenges in predicting a few complex OOD test clusters, causing significant performance variations. We also found that CGCNN, ALIGNN, and DeeperGATGNN performed more robustly on all OOD problems. By delving into the physical latent spaces of the trained models, we identified possible reasons for their comparatively better OOD performance than the current best models in the MatBench leaderboard - coGN and coNGN. Our work laid a solid foundation for advancing GNNs in OOD materials property prediction with multiple open research directions. One obvious direction is designing a robust GNN algorithm that can combine key contributing features from the architectures of CGCNN, ALIGNN, and DeeperGATGNN. Incorporating OOD data handling methods like domain adaptation~\cite{matda} can be another promising endeavor in improving the OOD property prediction performances of current GNNs. One final research direction is to investigate different OOD test set generation methods, especially focusing on the test clusters that caused high prediction variances, and to study how to improve performance over those particular clusters by examining the physical significance of each target generation method.

\section{Methods} \label{sec:methods}

\subsection{State-of-the-art (SOTA) algorithms for structure-based material property prediction} \label{subsec:sota_gnns}

We have chosen to evaluate the OOD performance for the following top structure-based material property algorithms as reported in the MatBench study \cite{dunn2020benchmarking}. They are all based on graph neural networks (GNNs) with different properties.

\subsubsection*{CGCNN}
CGCNN, proposed by Xie and Grossman~\cite{cgcnn}, is the earliest known GNN for the materials property prediction problem. After converting the crystals into crystal graphs and other preprocessing steps, CGCNN serially applies $N$ graph convolutional layers and $L_1$ hidden layers to the input crystal graph which results in a new graph with each node representing the local environment of each
atom. Following the pooling operation, a vector representing the entire crystal is linked to $L_2$ hidden layers and subsequently connected to the output layer to generate predictions. 

The $l$-th convolutional layer updates the node feature of the $i$-th atom $v_i$ through a process of convolution involving neighboring atoms and bonds of atom $i$ using a nonlinear graph convolution function as given below:
\begin{equation}\label{eq:1}
    v_i^{(l+1)} = \phi(v_i^{(l)}, v_j^{(l)}, e_{{(i, j)}_k})
\end{equation}

In Eq.~\ref{eq:1}, $e_{{(i, j)}_k}$ denotes the edge feature of the $k$-th bond connecting atom $i$ and atom $j$, and $\phi$ denotes the convolution operator.

\subsubsection*{MEGNet}
MEGNet (Chen et al.~\cite{megnet}) first performs the preprocessing steps to convert the input into graph embedding consisting of node and edge vectors. After that, $N$ MEGNet layers are applied, which include two dense layers, followed by the graph convolution operation. Next, a readout method is applied to combine sets of atomic and bond vectors into a single vector, followed by several size-reducing dense layers to finally produce the single-valued prediction.

The convolution operator can be defined as follows:
\begin{align}\label{eq:2}
    e_{i, j}^{\prime} = \phi_e(v_i \oplus v_j \oplus e_{i, j})\\\label{eq:3}
    v_i^{\prime} = \phi_v((\frac{1}{\mathcal{N}(i)} \sum_{j\in \mathcal{N}(i)} e_{i, j})\oplus v_i)
\end{align}

In Eq.~\ref{eq:2}, and Eq.~\ref{eq:3}, $v_i$ denotes the node representation for node $i$, $e_{i, j}$ denotes the edge representation between node $i$, and $j$, $v_i^{\prime}$, and $e_{i, j}^{\prime}$ denotes the updated node representation and edge representation, respectively, $\mathcal{N}_{i}$ denotes node $i$'s neighborhood, $\phi_e$, and $\phi_v$ denote the edge update function, and the node update function, respectively, and $\oplus$ denotes the concatenation operator.

\subsubsection*{SchNet}
Schütt et al.~\cite{schnet} developed SchNet for molecules which can also be applied to crystalline solids. It first creates the embeddings for graphs from the input materials and then applies $N$ interaction blocks to it, which includes the graph convolutions operation. After that, an atom-wise (a recurring building block applied separately to the node vectors) layer (reduces feature size), and a shifted softplus operation is applied. The final output is generated after applying another size-reducing atom-wise layer and a sum pooling operation.

The convolution operator can be defined as follows:
\begin{equation}\label{eq:4}
    v_i^{\prime} = \sum_{j\in \mathcal{N}(i)} v_j \circ \phi(e_{i, j})
\end{equation}

In Eq.~\ref{eq:4}, $v_i$ denotes the node representation for node $i$, $e_{i, j}$ denotes the edge representation between node $i$, and $j$, $v_i^{\prime}$, $\mathcal{N}_{i}$ denotes node $i$'s neighborhood, and $\phi$ denotes the convolution operator.

\subsubsection*{DimeNet++}
DimeNet++, developed by Gasteiger et al.~\cite{dimenet++} is a faster and improved version of the previously proposed DimeNet~\cite{dimenet} primarily for molecular property prediction. DimeNet++ takes a different approach than traditional GNNs for this task by embedding and updating the messages between atoms ($m_{ji}$). This allows DimeNet++ to incorporate directional information, considering bond angles ($\alpha_{(kj, ji)}$), in addition to interatomic distances $d_{ji}$. DimeNet++ goes further by jointly embedding distances and angles using a spherical 2D Fourier-Bessel basis. The following equation updates messages between atoms $m_{ji}$:

\begin{equation}\label{eq:dim1}
    m_{ji}^{(l + 1)} = \phi(m_{ji}^{(l)}, \sum_{k \in \mathcal{N}_j \textbackslash \{i\}} f_{int}(m_{kj}^{(l)}, e_{RBF}^{(ji)}, a_{SBF}^{(kj, ji)}))
\end{equation}

In Eq.~\ref{eq:dim1}, $\mathcal{N}_{i}$ denotes node $i$'s neighborhood, $f_{int}$ denotes the interaction function, $e_{RBF}^{(ji)}$ denotes the radial basis function representation of $d_{ji}$, and $a_{SBF}^{(kj, ji)}$ denotes the spherical basis function representation of $d_{kj}$ and $\alpha_{(kj, ji)}$.

\subsubsection*{ALIGNN}
In the prepossessing step, ALIGNN (Choudhary and DeCost~\cite{alignn}) converts a crystal to a crystal graph as done in CGCNN, and calculates node and edge features and other required processing. Moreover, it creates a line graph of the original graph to incorporate the angle feature between bonds. ALIGNN first applies edge-gated graph convolution on the line graph, and utilizes the edge representation and triplet representation features concerning the angle between the edge pairs) from layer $l$ to update the triplet representation, and bond messages of layer $l + 1$. The updated bond messages from layer $l + 1$ are passed to the next stage where they are incorporated with the original graph and the node representation from layer $l$. Then through a second edge-gated graph convolution, the node and the edge representation of layer $l + 1$ are calculated. The equations for updating the node and the edge features are given below:

\begin{align}\label{eq:ag1}
    u^{(l + 1)}, t^{(l + 1)} &= \phi_{eg} (L(G), e^{(l)}, t^{(l)})\\\label{eq:ag1}
    v^{(l + 1)}, e^{(l + 1)} &= \phi_{eg} (G, v^{(l)}, u^{(l + 1)})
\end{align}

In Eq.~\ref{eq:1}, $u^{(l)}$, $t^{(l)}$, $v^{(l)}$, and $e^{(l)}$ denotes the bond message representation, triplet representation, node representation, and edge representation, respectively, of layer $l$, and $\phi_{eg}$ denotes the edge-gated graph convolution operator.

\subsubsection*{DeeperGATGNN}
Omee et al.~\cite{deepergatgnn} developed the global attention-based GNN model DeeperGATGNN which essentially overcame the over-smoothing issue of GNNs (where with the increase of graph convolution layers, all the node feature vectors of the graph eventually update to the same vector) with the inclusion of differentiable group normalization (DGN)~\cite{zhou2020towards} and skip-connections~\cite{he2016deep}, and can go beyond > 50 layers. The process begins with an initial graph-encoded material serving as the input. Following this, multiple Augmented Graph Attention (AGAT) layers, each containing 64 neurons, and a DGN (Dynamic Graph Network) are utilized. There is a skip-connection from the output of the $l$-th AGAT layer to the output of the $(l+1)$-th AGAT layer, implemented post-DGN application. Subsequently, a global attention layer is introduced, where the node feature vectors are merged with the composition encoded vector. These are then processed through two fully connected layers, resulting in a context vector that encapsulates weights associated with the positions of each node. This context vector is then combined with the node feature vectors, followed by a global pooling of these vectors. The node features undergo further processing through one or two hidden layers, and finally, the output property is generated through an additional fully connected layer.

The local soft-attention $\alpha_{i,j}$ between a node $i$ and its neighbor $j$ can be represented by the following rule:
\begin{equation}\label{eq:dg1}
    \alpha_{i,j} = \frac{\text{exp}(a_{i,j})}{\sum_{k \in \mathcal{N}_{i}}{\text{exp}(a_{i,k})}}
\end{equation}
In equation~\ref{eq:dg1}, $\mathcal{N}_{i}$ denotes the node $i$'s neighborhood, and $a_{i,j}$ denotes the weight coefficient between nodes $i$ and $j$, indicating the significance of node $j$ concerning node $i$. The global attention $g_i$, employed just before global pooling, computes the overall importance of each node. It can be expressed by the following equation:
\begin{equation}\label{eq:dg2}
    g_{i} = \frac{ (x_i\parallel E) \cdot W} {\sum_{x_c \in X}{(x_c \parallel E) \cdot W}}
\end{equation}
In equation~\ref{eq:dg2}, $x \in \mathbb{R}^{F}$ denotes a learned embedding, $E$ denotes a compositional vector of the crystal, $W \in \mathbb{R}^{1\times (F+|E|)}$ denotes a parameterized matrix, and $x_c$ denotes the learned embedding of any atom $c$ within the crystal. 

\subsubsection*{coGN and coNGN}
coGN and coNGN (Ruff et al.~\cite{cogn}) use the basic GNN framework of Battaglia et al.~\cite{battaglia2018relational}, where a single GNN layer is defined by a graph network (GN) block responsible for converting a generic graph with edge, node, and global graph attributes using three update functions $\phi$ and three aggregation functions $\rho$. For the original material encoded graph $G$, a line graph $L(G)$ is constructed, so that there is an edge $e_{e_{ij}, e_{jk}}^{L(G)}$ for every two incident edges $e_{ij}, e_{jk}$ in $G$ (referring to the angle information between those two edges). Each nested GN block (total $T$ such blocks are applied) takes the edge features $x_E$, node features $x_V$, graph level features $x_G$, and the encoded graph $G$ itself as the input and outputs the updated node representation $x_V^{\prime}$, updated edge representation $x_E^{\prime}$, updated graph level representation $x_G^{\prime}$, and the graph $G$. The edge, node, and graph level representation update operations of coGN are given below:

\begin{align}\label{eq:co1}
    x_{e_{ij}}^{\prime} &= \phi_E(x_{e_{ij}}, x_{v_i}, x_{v_j}, x_G)\\\label{eq:co2}
    x_{v_i}^{\prime} &= \phi_V(x_{v_i}, \hat{x}_{v_i} , x_G)\\\label{eq:co3}
    x_G^{\prime} &= \phi_G(x_G, \hat{x}_G, \Tilde{x}_G)
\end{align}

In Eq.~\ref{eq:co1}, \ref{eq:co2}, and \ref{eq:co3}, $x_{e_{ij}}$ denotes the edge representation of edge $e_{ij}$ between node $i$, and $j$, $x_{v_i}$ denotes the node representation of node $i$, $\hat{x}_{v_i}$ denotes the local edge aggregated representation of node $i$, $x_{v_i}^{\prime}$ denotes the updated node representation of node $i$, $\hat{x}_G$ denotes the node aggregated representation of graph $G$, and $ \Tilde{x}_G$ denotes the global edge aggregated representation of graph $G$.
In the nested version (coNGN), the edge update is further continued by incorporating the angle information ($x_{\angle}$) from the line graph $L(G)$ using the following equation:

\begin{equation}
    \_ , x_E^{\prime} , \_ , \_ = \texttt{GN-Block}(x_{\angle}, x_E^{\prime}, x_G, L(G))
\end{equation}

\subsection{Evaluation criterion}
We use the mean absolute error (MAE) metric, which is a standard evaluation criterion for regression based materials property prediction problems. MAE can be calculated by the following equation:

\begin{equation*}
\text{MAE} = \frac{1}{n} \sum_{i=1}^{n} \left| y_i - \hat{y}_i \right|    
\end{equation*}

where \(y_i\) denotes the ground true property values, \(\hat{y}_i\) denotes the predicted property values, and \(n\) denotes the number of data points in the dataset.

\section{Data and Code Availability}
Details about the datasets and source codes of the GNN algorithms can be found at \url{https://github.com/sadmanomee/OOD_Materials_Benchmark}.

\section{Contribution}
Conceptualization, J.H.; methodology, S.O., J.H.,  N.F.; investigation, S.O, J.H., M.H.; software, S.O., N.F.; writing--original draft preparation, S.O., N.F., R.D., J.H.; writing--review and editing, S.O., N.F., R.D., M.H., J.H.; visualization, S.O., N.F. ; supervision, J.H. and M.H.

\section*{Acknowledgement}
The research reported in this work was supported in part by National Science Foundation under the grant and 10011239,10013417, and 10013216. The views, perspectives, and content do not necessarily represent the official views of the NSF.

\bibliographystyle{unsrt}  
\bibliography{references}

\end{document}